\documentclass[epj]{svjour}
\usepackage{epsfig,amsfonts,amssymb,latexsym,amsmath,bm,color,array,epsfig,graphics}

\newcommand{\be}{\begin{equation}}
\newcommand{\ee}{\end{equation}}
\newcommand{\bea}{\begin{eqnarray}}
\newcommand{\eea}{\end{eqnarray}}
\newcommand{\vek}{{\bm k}}
\newcommand{\veP}{{\bm P}}
\newcommand{\ver}{{\bm r}}
\newcommand{\vey}{{\bm \rho}}
\newcommand{\GeV}{\text{GeV}}

\usepackage{units} 
\usepackage{subfig} 

\graphicspath{{Figs.dir/}}

\begin{document}

\title{Remarks on meson loop effects on quark models}

\author{I.K. Hammer\inst{1} \and C. Hanhart\inst{1} \and A.V. Nefediev\inst{2,3,4}}       

\institute{Forschungszentrum J\"ulich, Institute for Advanced Simulation, Institut f\"ur Kernphysik and
J\"ulich Center for Hadron Physics, D-52425 J\"ulich, Germany \and 
Institute for Theoretical and Experimental Physics, 117218, B.Cheremushkinskaya 25, Moscow, Russia \and
National Research Nuclear University MEPhI, 115409, Kashirskoe highway 31, Moscow, Russia \and
Moscow Institute of Physics and Technology, 141700, 9 Institutsky lane, Dolgoprudny, Moscow Region, Russia}

\abstract{
We investigate the effect of meson loops on the spectrum of quark states. We demonstrate that in general quark states do not tend to get very broad if 
their coupling to the continuum increases, but instead they decouple from the latter in the large coupling limit. We ascribe this  
effect to the essentially nonperturbative unitarization procedure involved. In the meantime, some quark resonances 
behave very 
differently and demonstrate collectivity in the sense that their pole trajectories span a wide, as compared to the level spacing, 
region therefore acquiring contributions from multiple bare poles rather than from the closest neighbors. While the actual calculations are done 
within particular, very simplified models, it is argued that the findings might well be general. %
} 

\date{}
\maketitle

\section{Introduction}

Quark models for strong interactions have a long history since the middle of the previous century when it was commonly accepted that hadrons 
were not elementary particles, and thus the idea of quarks as elementary building blocks was put forward. In its 
simplest form, quark models describe 
the confining interaction between effective constituent quarks which leads to the formation of hadrons as bound states. Such hadrons appear stable unless the 
model is ``unquenched'', that is, unless it incorporates a pair creation mechanism which enables strong decays. It was speculated 
then that,
because of the large phase space available, the states high up in the spectrum acquire large widths and therefore they cannot be 
observed. In addition, hadronic loops, which bring an imaginary contribution to the mass of the 
state usually interpreted as width, also change
the real part of the pole thus resulting in a hadronic shift---see, for example, Ref.~\cite{Barnes:2007xu} and references therein. In particular, in 
Ref.~\cite{Barnes:2007xu} a systematic approach to hadronic shifts was developed and a number of loop theorems was formulated and proved.
On the other hand, it has been argued for a long time that when building quark models with coupled channels, unitarization cannot be neglected and 
even the appearance of extra, dynamically generated resonances is possible---see, for example, 
\cite{vanBeveren:1979bd,Tornqvist:1995ay,Boglione:1997aw,vanBeveren:2003vs,vanBeveren:2006ua,vanBeveren:2008rs,Rupp:2015taa,Rupp:2012py,Wolkanowski:2015lsa}
and references therein. Furthermore, in a particular model, contrary to naive expectations, quark states may even become narrower as the coupling 
to the continuum grows~\cite{vanBeveren:2006ua,AEK}.

In this paper we have a fresh look at the  phenomenon of ``unquenching''. In particular, we discuss under which 
circumstances it is appropriate to treat the coupling to the continuum perturbatively and when large corrections from unitarization are to be 
expected. In this context it turns out that the real part of the hadronic loops plays a crucial role for the behavior of the pole trajectories.
As these loops are divergent in an effective field theory their values must be fixed by some renormalization conditions that allow
one to relate the strength of the real parts to observables. In the study we want to pursue here, aimed at an understanding
of gross features that result from unitarization and not at a proper description of experimental data, we do not have such an option. 
Therefore we simply study the effect of different values for the real parts of the loops. Phenomenologically one may interpret
this variation as studying different models, since each quark model generates its own vertex functions that in turn lead to
specific real parts of hadronic loops. This will become evident from the discussion in Sec.~\ref{sec:ExpDecreasingCouplings}.

While already in the famous work \cite{Eichten:1978tg} and various later studies---for recent ones
in the heavy quark sector we refer to Refs.~\cite{Ortega:2010qq,Entem:2016ojz,Cincioglu:2016fkm}---unitarization effects are 
considered, to the best of our knowledge no detailed study of the effect of unitarization in the sense outlined above exists up to now.
The goal of this study is to investigate systematically the effect of the coupling of a tower of quark states to a continuum channel in a unitary 
framework.
For this we start from some model where the resonances are decoupled from the continuum. Then, the pole trajectories are followed as
the coupling to the continuum channel is switched on and increased.

While being of interest on its own, this research may be viewed as a naive study of the QCD spectrum as the 
number of colors, $N_c$, gets 
reduced from some very large value~\cite{'tHooft:1973jz,Witten:1980sp}
(for a pedagogical introduction to the large-$N_c$ limit of QCD we refer to Ref.~\cite{Donoghue:1992dd}).
In the limit $N_c\to \infty$ the QCD spectrum comprises infinite towers of stable (with respect to strong interactions) hadrons, since the 
strength
of three-point vertices scales as $1/\sqrt{N_c}$. At the same time the leading-order in 
$N_c$ contribution to the mass of $\bar{q}q$ states 
turns out to be independent of the number of colors---fully in line with the general features of the model employed in 
this paper.\footnote{As explained
in detail \emph{e.g.} in Ref.~\cite{Pelaez:2015qba}, the large-$N_c$ limit allows for deep insides into the nature of hadronic states. The approach 
we take here is complementary to that of Ref.~\cite{Pelaez:2015qba}: while there the number of colors is increased from the starting point $N_c=3$,
here we model the effect of decreasing $N_c$ from some very large value, admittedly with a much lower degree of rigor.}

\section{Concept of modeling, important results and disclaimers}

\begin{figure}
\centering
\includegraphics[width=.45\textwidth]{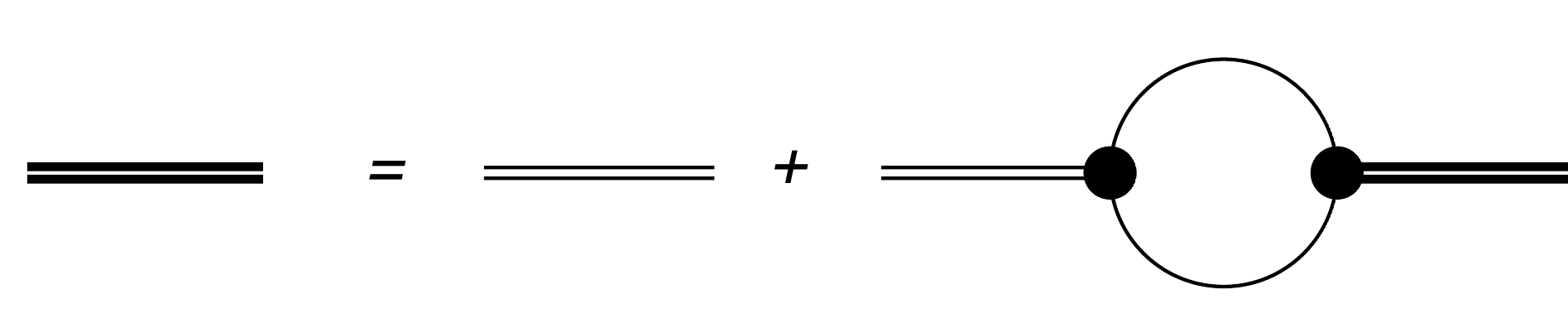}\\
\caption{Illustration of the Dyson equation: The physical Green's function is found from
coupling individual quark states via meson loops. The thick double line denotes
the physical Green's function while the thin double line denotes the bare
quark model states. Single lines denote the particles of the continuum channel.}
\label{fig:Dyson}
\end{figure}

The main purpose of this research is to investigate in general terms the effect of hadronic 
loops on a spectrum of quark states. Thus, the models employed below are not intended to be used for realistic calculations 
of hadron properties as they are only meant to capture the gross features dictated by QCD.
To be concrete, for the confinement potential we explicitly studied a linear as well as a 
quadratically rising potential. The resulting states are included in the $s$ channel only.
While performing each calculation with a fixed number of quark model states, $N$, we confirm that
the general features of the pole trajectories
do not depend on the number of basis states by repeating the calculation for different values of $N$, that even allows
us to speculate about the $N\to \infty$ limit. 

In this work 
the mentioned quark model states get coupled to each other via a meson loop in a single continuum channel. For the 
structure of 
the corresponding vertex functions again we use two different, simple models. The resulting Dyson equation for the physical Green's function is shown
diagrammatically in Fig.~\ref{fig:Dyson}. 

In our study the overall coupling strength for each resonance to the continuum channel is controlled
by a single parameter that is varied for all quark states simultaneously. This procedure leads to certain pole 
trajectories, as a function
of the mentioned strength parameter, that are studied for the different models mentioned above.
For definiteness, the mass parameters in these calculations are chosen in a range that corresponds to $\bar{c}c$ states 
coupled to a pair of open-charm mesons. 
The most important findings of this work are:
\begin{itemize}
\item[$\bullet$] For small couplings each individual state located above a relevant threshold 
acquires a width that scales as the square of the coupling. The 
other states present in the system do not affect this behavior.
\item[$\bullet$] As the coupling increases the width does not keep on growing for most of the states but, as soon as the
selfenergies get of the order of the level spacing, the pole trajectories bend and the widths decrease
again. A typical shift of the real part of such a pole compared to its bare value is of the order of the level spacing, and 
the pole is therefore mostly influenced by the nearest neighbors.
\item[$\bullet$] Besides the states just described at least one state in the spectrum behaves very differently. As the coupling increases, the 
trajectory of the corresponding pole is not bounded by the nearest neighbors but it covers several level spacings thus 
going far away from the 
original, bare pole position. As a result it is sensitive not only to its neighbors but also to remote states. For 
this reason we refer to 
such poles as to collective ones. 
\item[$\bullet$]  Which state demonstrates this collective phenomenon and if this collectivity shows up in one or more states depends
on the renormalization condition imposed and on the details of the model (type of the vertex function, number of states
included, and so on).
\end{itemize}

Thus we found that in general, as soon as the coupling of quark states to the continuum gets large, 
unitarization effects cannot be neglected and they change severely the properties of the states. 
In particular, for large couplings there appear two types of states: 
ordinary and extraordinary states---this notion was introduced by Jaffe in 
Ref.~\cite{Jaffe:2007id}---which behave quite similarly for small couplings and are therefore indistinguishable in the 
weak coupling regime. 
Meanwhile, they behave quite differently as the coupling increases. In particular, for large couplings,
the ordinary states tend to decouple from the continuum. Each of those states contains a prominent contribution
from the original basis state and the admixture from other states decreases quickly with their distance
to the pole. In the same limit, the extraordinary states demonstrate a collective phenomenon in the 
sense of sensitivity to multiple bare states, and it is this collectivity that could 
give us a clue on their physics as discussed in Sec.~\ref{sec:int}.

\begin{figure}
\centering
\includegraphics[width=.15\textwidth]{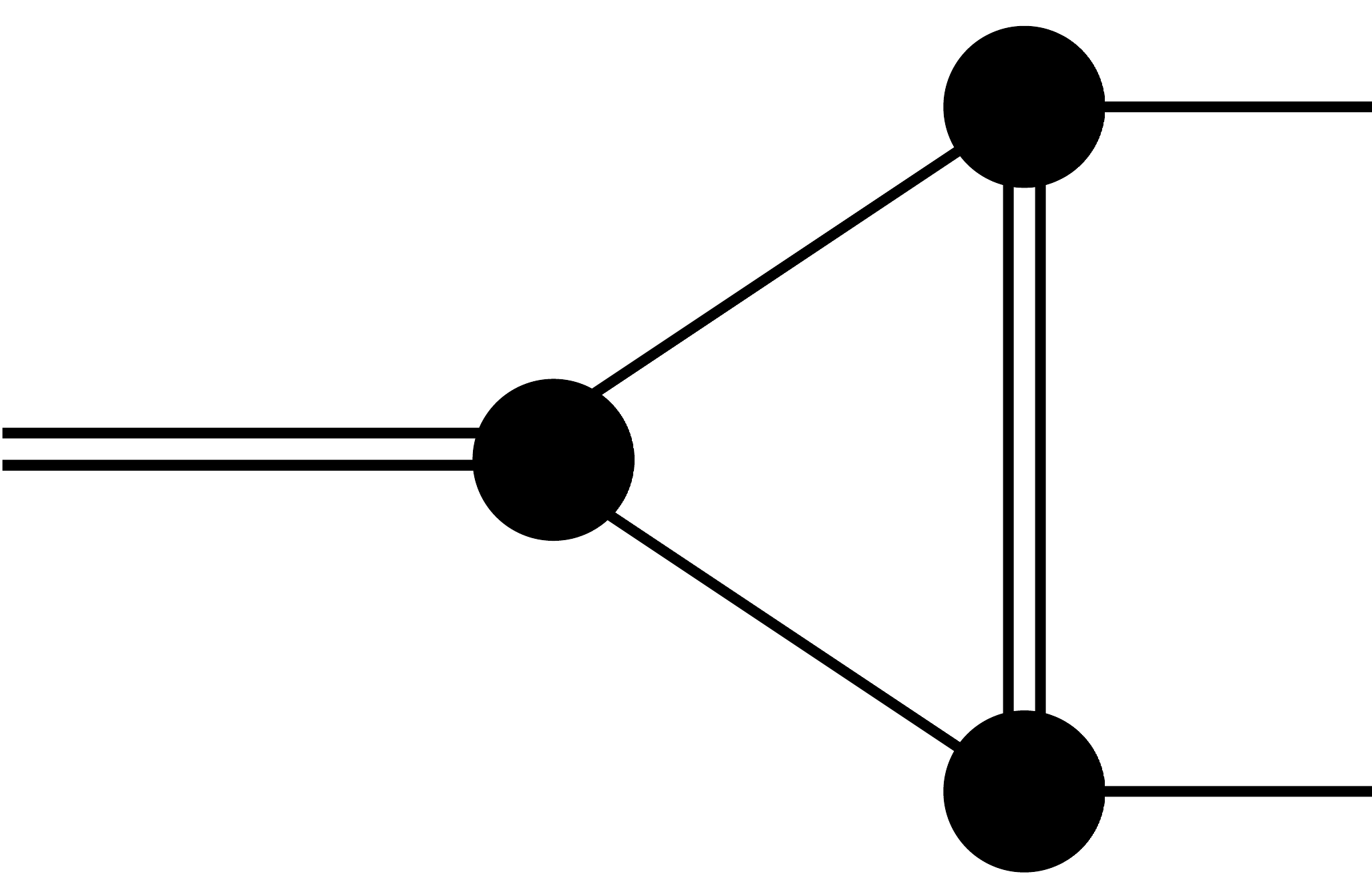}\\
\caption{Possible vertex dressing diagram $not$ included in the
present study. The meaning of the lines is the same as in Fig.~\ref{fig:Dyson}.}
\label{fig:vertex}
\end{figure}

It should be stressed that the models we used are very simplified and by no means 
realistic. For example, we do not
include $t$-channel exchanges of heavy mesons in the vertex dressing (as indicated in Fig.~\ref{fig:vertex}) --- although
formally they should be included ---
but just restrict ourselves to simplified models for the vertices whose structures are independent of the 
coupling strength. As was demonstrated in Ref.~\cite{Schneitzer:2014rsa}, the contribution of the triangle diagram from 
Fig.~\ref{fig:vertex} is indeed suppressed in the weak-coupling regime ($g\to 0$). However,
since the focus of this work is on qualitative studies of the behavior of the pole trajectories we dare neglect this diagram also for relatively large values of 
the coupling. A  more refined study of the role played by the triangle diagram is left for future publications.
We also do not allow for a $t$-channel exchange of light mesons in this work although the left-hand cuts
they induce could lead to additional structure: while those will certainly change the quantitative behavior
of the trajectories we do not expect the gross features to be changed.
In addition, we only use a single continuum channel as well as a
simplified form for the two-meson loop whose real part is controlled by just a single 
counter term. Still, regardless of those severe simplifications, since 
the two aforementioned types of states appear for all model versions we studied, we are tempted
to claim the mentioned features to be general and therefore very likely to reveal themselves also in more realistic models. 
More refined studies will be presented in a subsequent publication.

\section{Some instructive Examples}
\label{sec:Some instructive Examples}

We start our presentation with a short discussion of systems with only one or two resonances where some features of the unitarization and
regularization can already
be discussed using simplified analytic calculations. In the subsequent Sections we turn to more general cases.

For simplicity we consider only one continuum channel, so that the Riemann surface consists of two sheets: the first (physical) sheet and the 
second (unphysical) sheet. Poles on the physical sheet are only allowed to reside on the real axis below the threshold and they correspond to
bound states. The poles on the unphysical sheet located on the real axis below the threshold correspond to virtual states. All other poles on the 
second 
sheet correspond to resonances.

\subsection{Single-resonance case}

As the simplest example, we consider a single scalar resonance $R$ coupled to a continuum channel $\varphi\bar{\varphi}$ with 
$\varphi$($\bar{\varphi}$) being a scalar (anti)meson, 
\begin{eqnarray}\nonumber
{\cal L}=\frac12\left[(\partial_\mu\varphi)^2-m^2\varphi^2\right]
+\frac12\left[(\partial_\mu\bar{\varphi})^2-m^2\bar{\varphi}^2\right] \\
+\frac12(\partial_\mu R)^2-\frac12 M^2R^2+gR\bar{\varphi}\varphi.
\label{eq:Lagrangian1}
\end{eqnarray}

The physical propagator of the resonance is 
calculated from the Dyson equation $G=G_0-G_0\Sigma G$ with the solution
\begin{eqnarray}
 G=\frac{G_0}{1+G_0\Sigma}=\frac{1}{s-M^2+\Sigma}\ ,
\label{G1res}
\end{eqnarray}
where $G_0=1/(s-M^2+i0)$ is the bare scalar propagator, $\Sigma$ is the selfenergy,
\begin{equation}
\Sigma =-g^2\Pi,
\end{equation}
$g$ is a real coupling constant, and $\Pi=\frac1i\raisebox{-1.7mm}{\epsfig{file=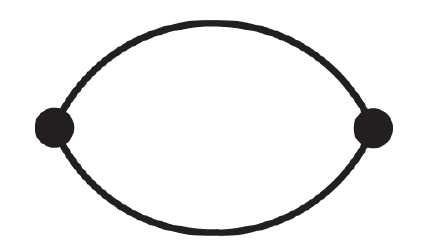,width=9mm}}$ is the 
loop operator. If $\Pi$ is 
treated as an energy-independent complex number then the pole of the dressed propagator (\ref{G1res}) is located at $s_p=M^2+g^2\Pi$. If now
$\mbox{Im}\Pi\neq 0$ then 
$|\mbox{Im}(s_p)|\rightarrow \infty$ in the limit $g\to\infty$ that means that, for a constant selfenergy the width 
grows for an increasing coupling and it can become arbitrarily large. 

In contrast to the toy example above, a more realistic selfenergy $\Sigma$ is $s$-dependent. This is a necessary consequence 
of the analytic 
properties of the $S$ matrix that require the presence of a pole at $s_p^*$ if there is a pole at $s_p$. To proceed we 
need to specify the continuum 
channel for which we stick to the simplest two-body system formed by a scalar particle $\varphi$ and its antiparticle 
$\bar{\varphi}$ which can 
interact only via the resonance $R$. To leading order in the center-of-mass momentum 
\be
k=\frac12\sqrt{s-4m^2},
\label{ks}
\ee
where $m$ is the mass of the field $\varphi$, the selfenergy operator reads
\begin{equation}
\Pi(s)=-\frac{i}{16\pi m}k-\Delta(g).
\label{piofs}
\end{equation}

The term $\Delta(g)$ is a real function which depends only on the masses and on the coupling constant $g$. 
The function $\Pi(s)$ is supposed to parametrize the leading energy dependence of the underlying loop function
depicted in Fig.~\ref{fig:Dyson}. Since this loop is divergent, as long as no form factors are introduced, some regularization
procedure is required to fix its real part. 
Since in this exploratory study we cannot use experimental information to fix the real part of the loop,
we study the impact of different values for $\Delta(g)$. Phenomenologically this means that we simulate in this way
different quark models, since those will generate different hadronic form factors that in turn lead to different values
for the real part of $\Pi(s)$. 
We come back to this discussion in Sec.~\ref{sec:ExpDecreasingCouplings}.

Note that unitarity requires the spectral density
\be
w(s)=-\frac{1}{\pi}\mbox{Im}G(s)
\label{w}
\ee
to be normalized as
\be
{\cal N}=\int_{s_{\rm th}^{\rm min}}^\infty w(s)ds=1-\sum_n Z_n,
\label{norm}
\ee
where $s_{\rm th}^{\rm min}=(2m)^2$ is the lowest open threshold 
and the sum runs over all bound states (first sheet poles). Here $Z_n$ denotes
the corresponding wave function renormalization constants that may be interpreted as
 the probability to find the $n$-th bare state in the continuum wave 
function \cite{Bogdanova:1991zz,Baru:2003qq}.\footnote{For a recent discussion see, for example, Ref.~\cite{Giacosa:2007bn}.} It is easy to verify 
that the spectral density constructed with the help 
of Eq.~(\ref{G1res}) is indeed normalized to unity for small couplings $g$ for all values of the subtraction constant 
$\Delta$. 
Meanwhile, as the coupling grows, for a particular sign of $\Delta$, it
happens that additional poles enter the physical Riemann sheet and turn into bound states. 
In this case the spectral density (\ref{w}) is not normalized to unity any more, and the deviation of the normalization integral from unity
is given by Eq.~(\ref{norm}) with nonzero $Z$'s. As before, this feature is independent of a particular choice of 
$\Delta$.

Substituting function (\ref{piofs}) into Eq.~(\ref{G1res}) it is straightforward to find the poles of 
the physical propagator,
\begin{eqnarray}\nonumber
s_p&=&M^2-g^2\Delta-\frac{1}{8}\left(\frac{g^2}{16 \pi m}\right)^2\\
&\pm&\frac{g^2}{32 \pi 
m}\sqrt{4m^2{-}M^2{+}
\frac{1}{8}\left(\frac{g^2}{16 \pi m}\right)^2+g^2 \Delta}\ ,
\label{sp1}
\end{eqnarray}
which depend on $\Delta$. 

As the first example we choose as renormalization condition the requirement that the real part of the pole studied does not move
when the coupling to the continuum is switched on.
This leads to the condition
\begin{equation}
\Delta(g)=-\frac{1}{8}\frac{g^2}{(16\pi m)^2},
\end{equation}
for which Eq.~(\ref{sp1}) for the poles position simplifies considerably to read
\begin{equation}
s_p=M^2\pm\frac{1}{2}\left(\frac{g^2}{16\pi m}\right)\sqrt{4m^2-M^2}.
\label{sp1b}
\end{equation}
For $M>2m$ the square root in Eq.~(\ref{sp1b}) is always imaginary and the width of the 
state $R$ grows quadratically with $g$. This behavior is 
similar to the one for 
the constant $\Sigma$ considered above, however now the unitarity condition $s_p^{(2)}=\left(s_p^{(1)}\right)^*$ is 
fulfilled automatically.

Alternatively we could have chosen the condition $\Delta(g)$ $\equiv 0$, which implies that $\Delta$ is naively absorbed into a 
redefinition of the mass $M$. 
Then the poles position in the complex momentum plane---derived from Eq.~(\ref{sp1})---reads
\begin{eqnarray}\nonumber
k_p&=&-\frac{i}{8}\left(\frac{g^2}{16\pi m}\right)
\\
& & \qquad \pm\frac12\sqrt{M^2-4m^2-\frac{1}{16}\left(\frac{g^2}{16 \pi 
m}\right)^2}.
\label{eq:Solution_One_Resonance_Nonrel}
\end{eqnarray}

\begin{figure*}
\centering
\subfloat[]{\includegraphics[width=.35\textwidth]{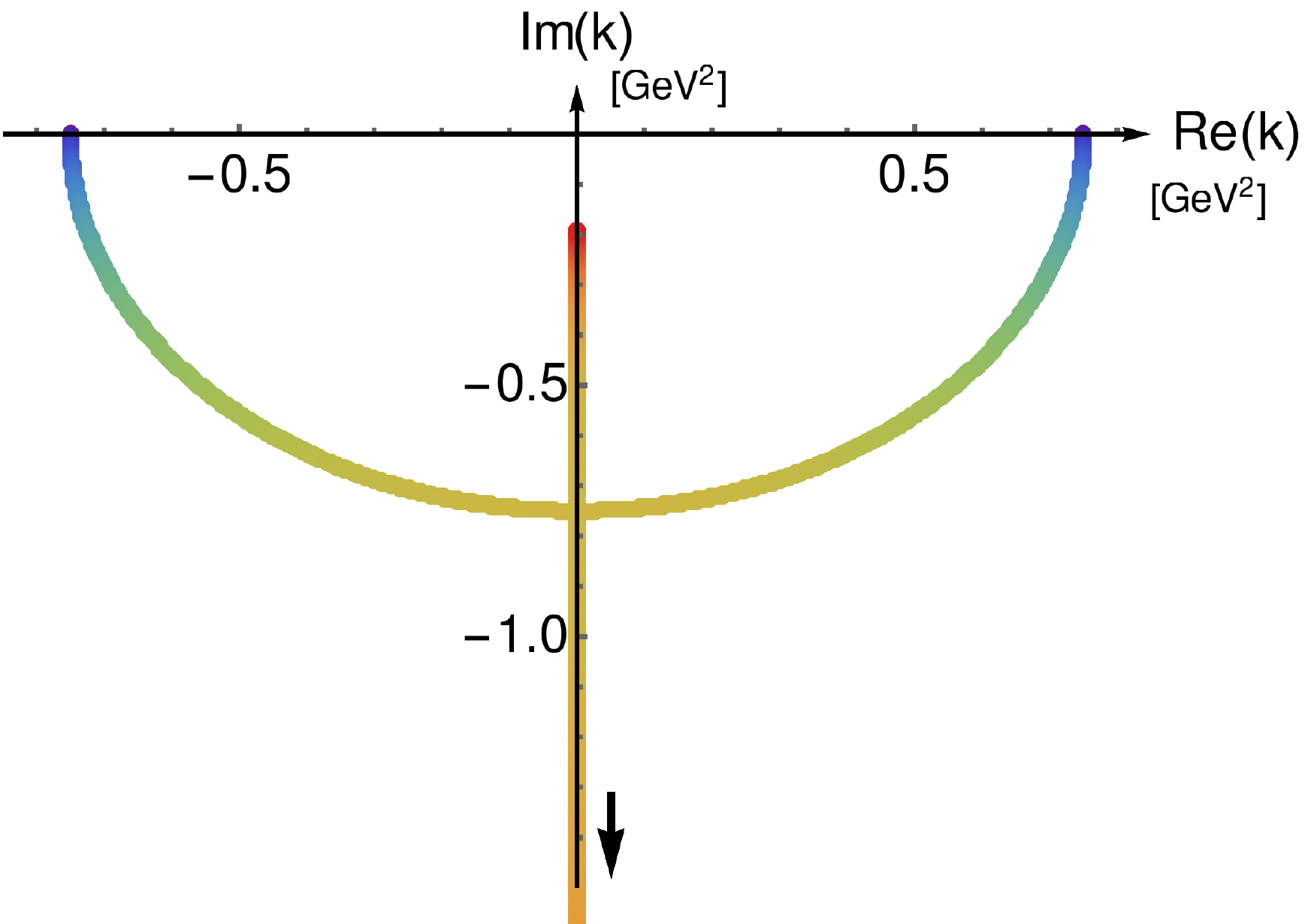}} \qquad
\subfloat[]{\includegraphics[width=.35\textwidth]{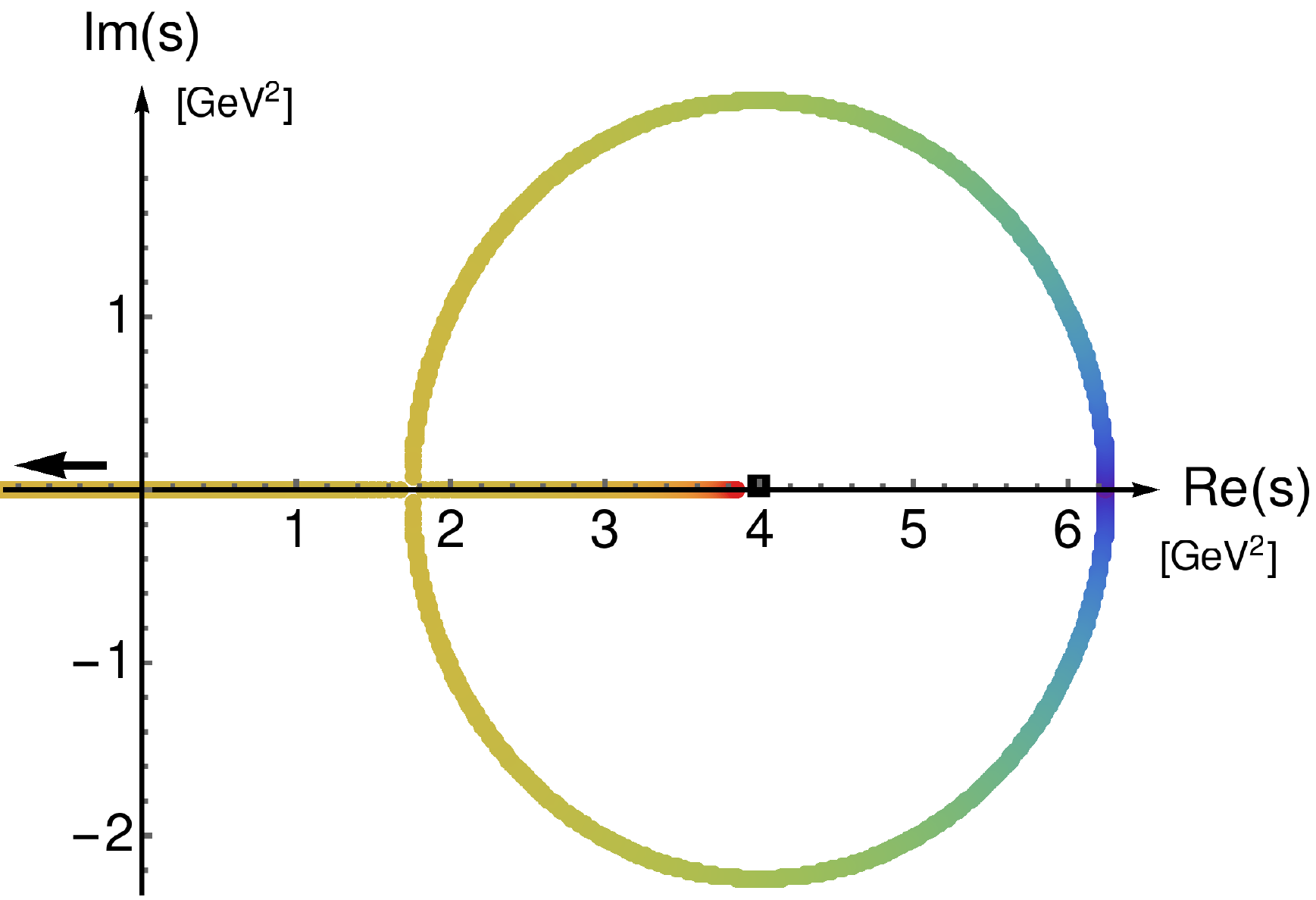}} \qquad
\includegraphics[width=.08\textwidth]{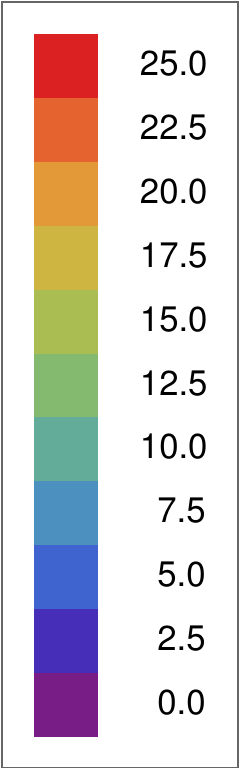}\\
\caption{Pole trajectories for the renormalization condition $\Delta(g)=0$ and for the values of $g$ between $0$ and 
$25$~GeV 
($M=2.5$~GeV, $m=1$~GeV). The left (right) panel shows the pole trajectories in $k$-plane ($s$-plane). The black square 
indicates the threshold.}
\label{fig:1StLoop}
\end{figure*}

In Fig.~\ref{fig:1StLoop}(a) we show the behavior of the poles $k_p$ as a function of the coupling
$g$ for $M>2m$. In the limit $g=0$ there are two 
symmetric 
real poles which describe a stable quark state $R$. 
As $g$ deviates from zero but remains small, $g^2\ll 64\pi m\sqrt{M^2-4m^2}$, the state 
$R$ acquires a small width and turns to a resonance. As $g$ increases further, contrary to the naive expectations, the 
width does not grow 
infinitely, but the poles approach each other, they collide at the imaginary axis for $g^2=64\pi m\sqrt{M^2-4m^2}$, and 
then, for $g\to\infty$, one 
pole leaves the near-threshold region while the other one approaches the threshold from below.
It is interesting to note that such asymmetrically located poles for near threshold states can be interpreted as a 
signature of a 
predominantly molecular nature of the state~\cite{Morgan:1992ge,Tornqvist:1994ji}. The underlying pole counting approach 
is in line with the famous compositeness criterion by Weinberg~\cite{Weinberg:1962hj}, as shown in 
Ref.~\cite{Baru:2003qq}.

The corresponding trajectories of the poles in the complex $s$-plane are shown in 
Fig.~\ref{fig:1StLoop}(b).\footnote{For a 
discussion of the pole trajectories from a different perspective we refer the reader to Ref.~\cite{Hanhart:2014ssa}. 
A detailed investigation of the interplay of quark and meson degrees of freedom in near-threshold resonances 
and the discussion of the resulting behavior of the poles can be found in 
refs.~\cite{Baru:2010ww,Hanhart:2011jz,Guo:2016bjq}. Coupled-channel dynamics of the poles describing near-threshold states can be found, for 
example, 
in Refs.~\cite{Kalashnikova:2005ui,Danilkin:2009hr,Danilkin:2010cc}.} 
In particular, for $g=0$ one has two degenerate poles lying on the real axis above threshold which then travel 
symmetrically in the second Riemann 
sheet until they collide again on the real axis below the threshold. Then, in agreement with the discussion of the 
poles in 
the complex $k$-plane, one of 
them leaves the near-threshold region while the other turns to a virtual state. It is easy to verify that the 
maximal imaginary part of the pole in the complex $s$-plane is $(M^2-4m^2)/4$. Therefore the state never gets 
arbitrarily broad and its maximal possible width is governed by the proximity of the bare poles to the threshold.

For $M<2m$ one can choose a different renormalization scheme with
\begin{equation}
\Delta=\frac{1}{32\pi m}\sqrt{4m^2-M^2},
\end{equation}
that ensures $\Sigma(s=M^2)=0$ for any $g$, so that the bound-state pole stays fixed at $s=M^2$. Meanwhile, its 
counterpart on the unphysical 
Riemann sheet moves away from the threshold with a growing value of $g$. Thus the system behaves
similarly to the previous case for large couplings.

\subsection{Two-resonance case}

\begin{figure*}
\centering
\subfloat[]{\includegraphics[width=.35\textwidth]{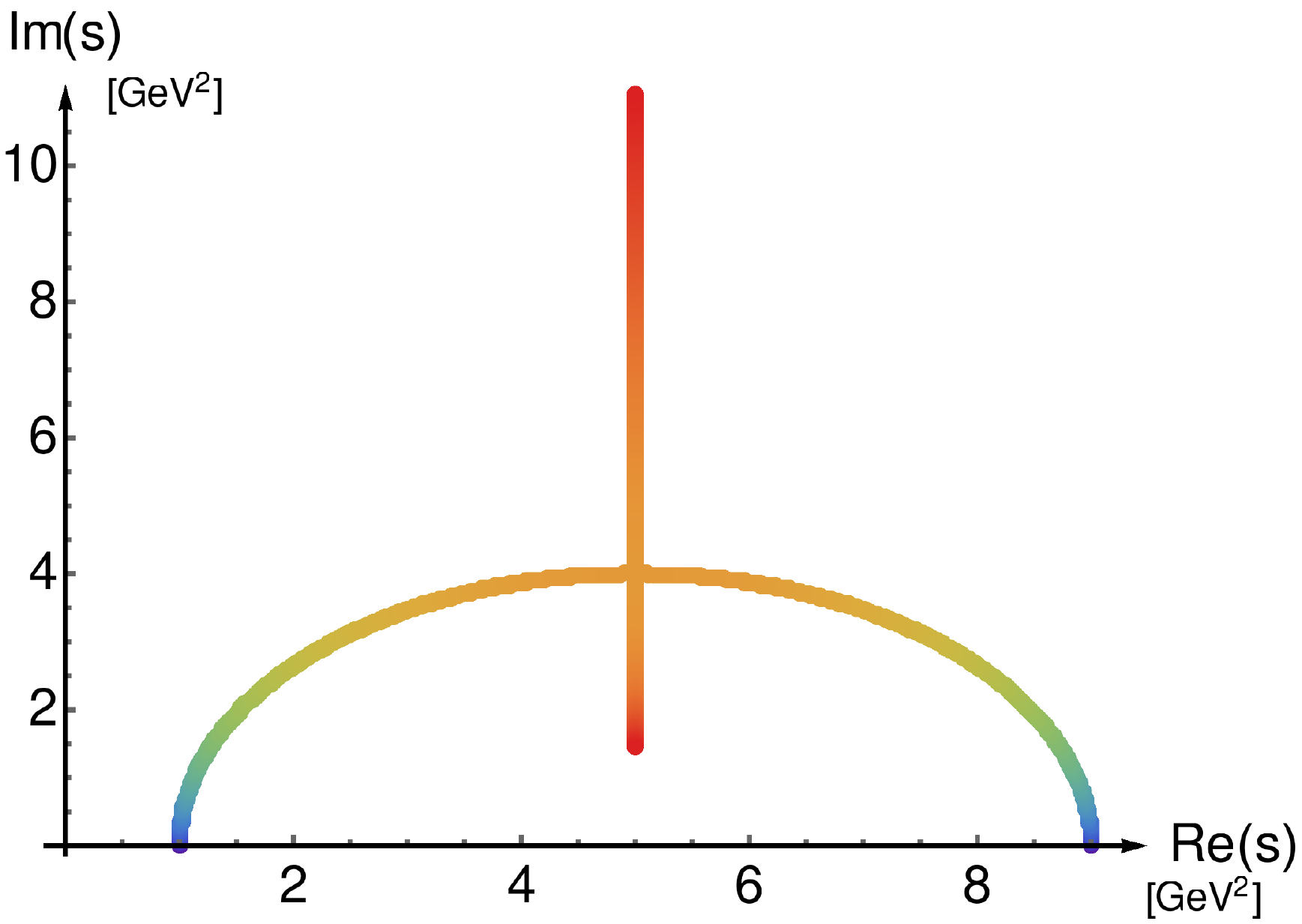}} \qquad
\subfloat[]{\includegraphics[width=.35\textwidth]{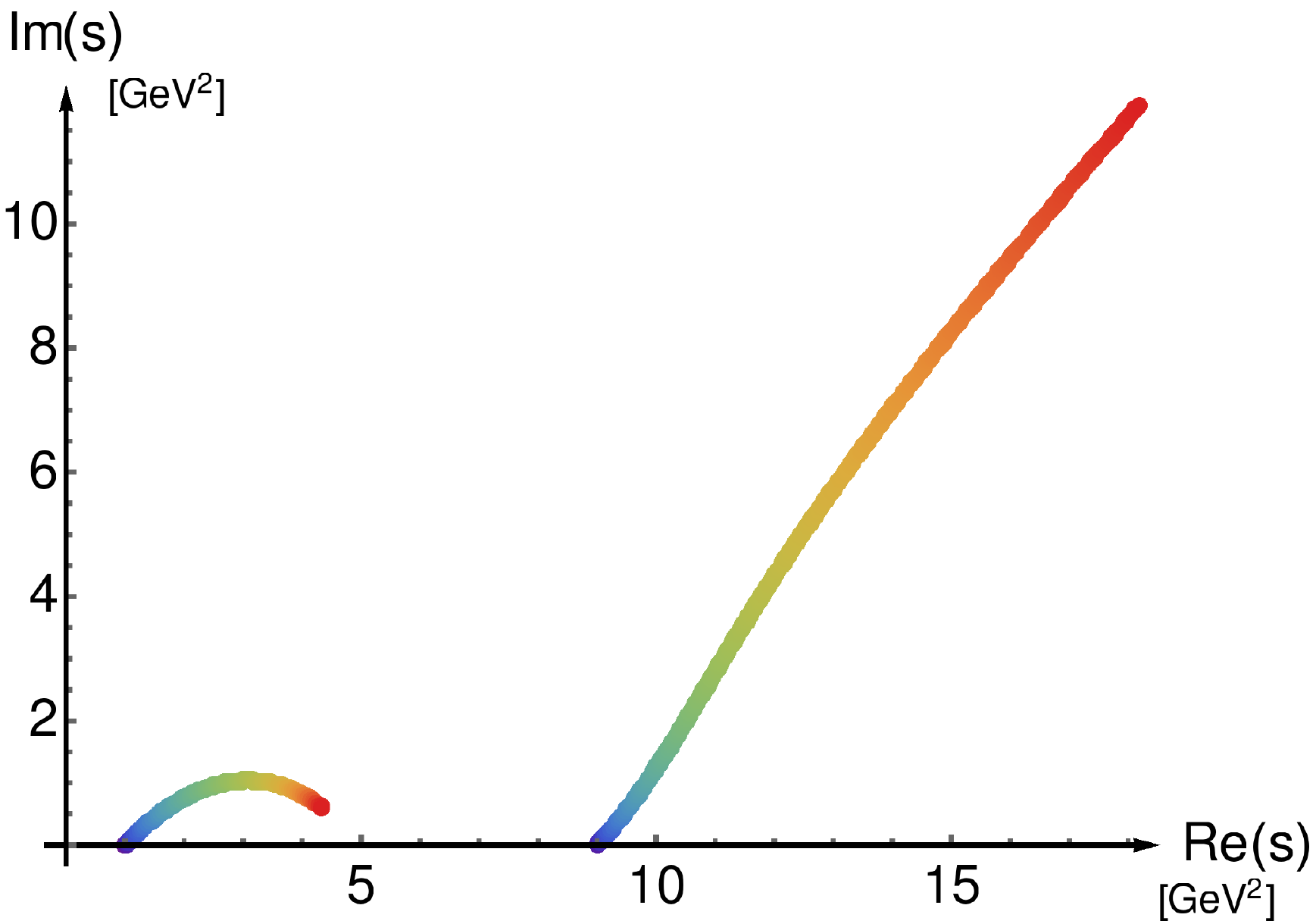}} \qquad
\includegraphics[width=.08\textwidth]{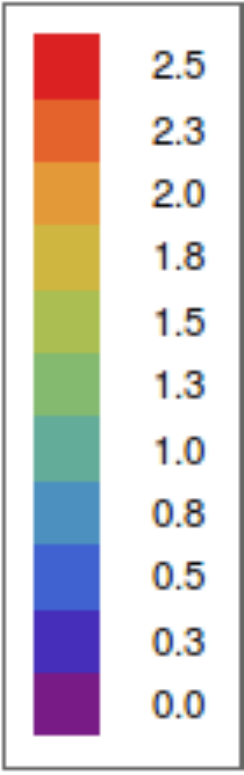}
\caption[2StConst]{Pole trajectories in the upper half of the complex $s$-plane as given in 
Eq.~(\ref{eq:TwoResonance_Result_for_s}) for $g$ growing 
from 0 to 2.5~GeV (see Eq.~(\ref{setparam2}) for the values of other parameters). The left(right) panel shows the result 
for $\Pi=i$($\Pi=1+i$).}
\label{fig:2StConst}
\end{figure*}

Now we consider a system of two resonances, $R_1$ and $R_2$, which, as before, couple to a single elastic 
channel $\varphi\bar{\varphi}$. The resonances therefore communicate through the off-diagonal elements of the selfenergy 
operator 
$$
\Sigma_{ij}=-g_i g_j\Pi.
$$
The poles of the scattering matrix in such a two-chanel problem come as solutions of the 
equation $\det\left(1+G_0\Sigma\right) = 0$, where $G_0=\mbox{diag}((s-M_1^2+i0)^{-1},(s-M_2^2+i0)^{-1})$. 
For definiteness, we fix the parameters of the system to values typical for a charmonium system, namely
\be
M_1=1~\GeV,\quad M_2=3~\GeV,\quad m=1~\GeV,
\label{setparam2}
\ee
that ensures that the elastic threshold is located between the bare resonances, $M_1<2m<M_2$. Also, for simplicity, we 
set $g=g_1=g_2$. 

If the loop operator $\Pi$ is assumed to be constant, the poles can be found analytically in the form
\begin{eqnarray}\nonumber
s_p&=&\frac12\left(M_1^2+M_2^2\right)+g^2\Pi
\\
& & \qquad \pm\sqrt{\frac{1}{4}\left(M_1^2-M_2^2\right)^2+\left(g^2\Pi\right)^2}.
\label{eq:TwoResonance_Result_for_s}
\end{eqnarray}

As before, we are interested in the behavior of the poles as the coupling $g$ grows. The picture becomes especially 
simple, if $\Pi$ is purely 
imaginary (for definiteness we set $\Pi=i$). In this case and in the limit $g\rightarrow\infty$ the poles given in 
Eq.~(\ref{eq:TwoResonance_Result_for_s}) become
\begin{eqnarray}\nonumber
s_p^{(1)}&=&\frac{1}{2}\left(M_1^2+M_2^2\right)+2ig^2\quad\mbox{and}\quad  \\
s_p^{(2)}&=&\frac{1}{2}\left(M_1^2+M_2^2\right)-\frac{i}{2g^2}\left(M_2^2-M_1^2\right)^2,
\end{eqnarray}
so that they demonstrate the pattern depicted in Fig.~\ref{fig:2StConst}(a): One pole acquires a large imaginary 
part and leaves the 
near-threshold region fast while the other pole gradually approaches the real axis, that is it becomes nearly stable. 
The relevant scale for its 
width is now given by the level spacing $M_2^2-M_1^2$. 

As the loop operator $\Pi$ acquires a real part (for illustration we set $\Pi=1+i$---see 
Fig.~\ref{fig:2StConst}(b)), the particular picture 
changes, however it still demonstrates the same pattern: as the coupling $g$ grows, one state gets stable again while 
the other one becomes very 
broad.

\begin{figure*}[t]
\centering
\includegraphics[width=.7\textwidth]{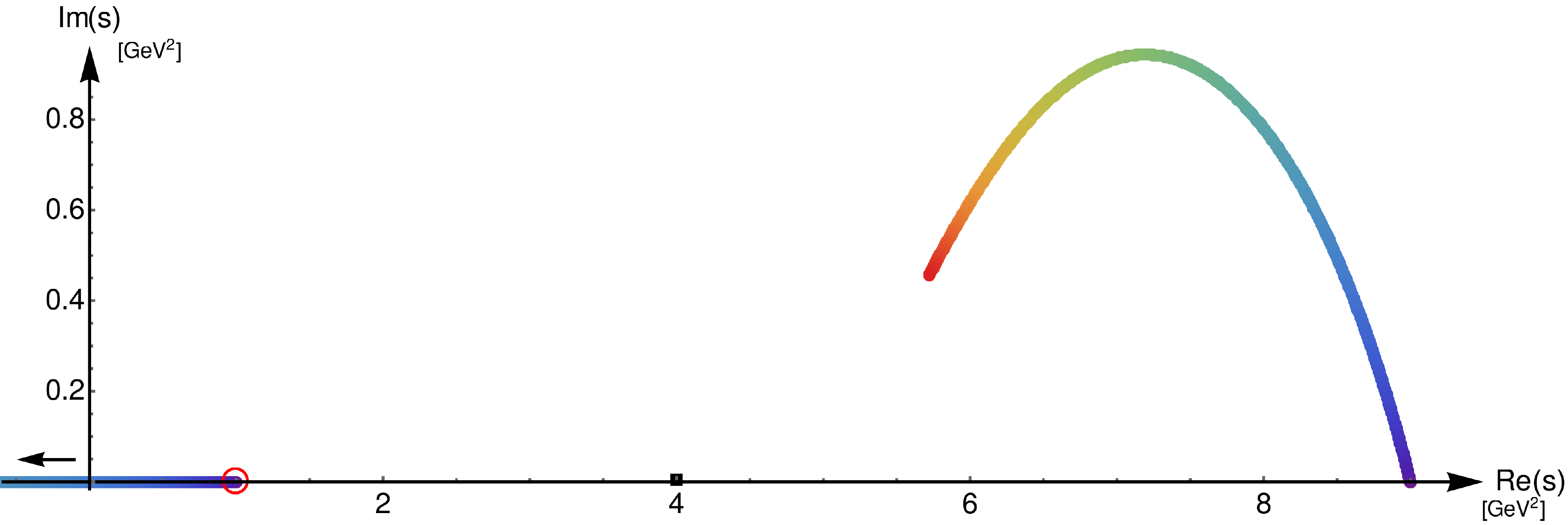} \quad
\includegraphics[width=.07\textwidth]{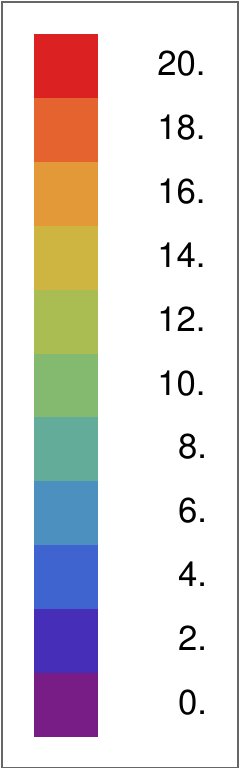}
\caption[2StLoopRen]{Pole trajectories in the upper half of the complex $s$-plane for the energy-dependent loop operator 
$\Pi(s)$ given by 
Eq.~(\ref{piofs}) with the $\Delta$ fixed as in Eq.~(\ref{Delta2ch}). The coupling $g$ is increased from 0 to 20. 
The black square indicates the threshold while the red circle shows the position of the bound state which is kept constant 
due to the particular choice of 
$\Delta$---see Eq.~(\ref{Delta2ch}).}
\label{fig:2StLoopRen}
\end{figure*}

Finally, we study the interplay of the different phenomena discussed in this Section. We consider the energy-dependent 
loop operator $\Pi(s)$ as given 
in Eq.~(\ref{piofs}) and fix the $\Delta$-term in the form
\be
\Delta=\frac{1}{32\pi m}\sqrt{4m^2-M_1^2} \ ,
\label{Delta2ch}
\ee 
which ensures that the position of the pole on the first Riemann sheet (the bound-state pole) remains on its location. 
At the same time its mirror pole on the second Riemann sheet moves away towards smaller and then negative values of
$s$. The results of the corresponding numerical calculations 
are shown in Fig.~\ref{fig:2StLoopRen}. In the meantime, the resonance pole (its mirrored counterpart is not shown in 
Fig.~\ref{fig:2StLoopRen}) acquires a small width as 
$g$ departs from 0. However, contrary to naive expectations, it does not become infinitely broad for large values of $g$. 
Instead, as $g$ exceeds some 
critical value the width tends to decrease, so that eventually, for $g\to\infty$, this state tends to decouple from 
the continuum.

Before we proceed to multi-resonance systems, let us briefly summarize the findings of this Section. On the one hand, using 
simple and transparent models 
we found that, contrary to naive expectations, at least one state does not become infinitely broad in the limit of a 
strong coupling 
to the hadronic channel---on the contrary, it even decouples from the continuum in this limit thus turning to an 
asymptotically stable 
object. In addition, we demonstrated that the renormalization condition for the hadronic loop is an essential ingredient of the 
model---the form of a particular pole trajectory and whether or not a particular state decouples from the continuum 
may depend strongly on it. Stated differently: Two different models which are able to describe the physical spectrum of 
had\-rons equally well may predict very different pole trajectories as soon as the coupling constants deviate from their 
physical values.

\section{Unitarization in multi-resonance systems}

In the previous Section we studied a single-resonance and a two-resonance system and found that the trajectories of 
the resonance poles that emerge as the coupling to the continuum is increased depend strongly on both the energy dependence of the selfenergy 
and on the renormalization condition for the loop. In this Section we consider multi-resonance systems and find the collective phenomena
already mentioned above.

\subsection{Multi-resonance toy model}

To study the coupling of a tower of quark states to a continuum channel, we consider a system governed by the 
Lagrangian 
\begin{eqnarray}\nonumber
{\cal L}=\frac12\left[(\partial_\mu\varphi)^2-m^2\varphi^2\right]
+\frac12\left[(\partial_\mu\bar{\varphi})^2-m^2\bar{\varphi}^2\right] \\
+\sum_{n=1}^N\left[\frac12(\partial_\mu R_n )^2-\frac12 M_n^2R_n^2+g_nR_n\bar{\varphi}\varphi\right],
\label{eq:Lagrangian}
\end{eqnarray}
which is a natural generalization of Lagrangian (\ref{eq:Lagrangian1}) to the multi-resonance case.
Here, as before, $\varphi$ and $R_n$ ($n \in \lbrace 1, \ldots, N \rbrace$) are scalar quark-antiquark fields, and 
$\bar{\varphi}$ is the charge 
conjugated 
field with respect to $\varphi$. The number of resonance states considered, $N$, is varied in what follows. It is natural to 
identify the 
fields $R_n$ with the $n$-th radial excitation of the $\bar{Q}Q$ meson, where $Q$ is a heavy quark. Then $\varphi$ and 
$\bar{\varphi}$ are
heavy-light $\bar{Q}q$ and $\bar{q}Q$ mesons, respectively, produced from $R_n$ through some string breaking mechanism. 

The $T$ matrix for $\varphi\bar{\varphi}$ scattering may be written as $T=(1+V\Pi(s))^{-1}V$, where the potential reads 
\begin{equation}
V=\sum_{n=1}^N\raisebox{-5mm}{\includegraphics[scale=1]{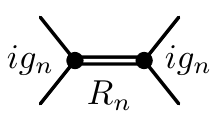}}=-\sum_{n=1}^N \frac{g_n^2}{s-M_n^2}.
\label{eq:_potential_original}
\end{equation}
For simplicity, for $\Pi(s)$ we again resort to the scalar loop up to the leading order in $k$, as defined in 
Eq.~(\ref{piofs}). For the masses of the particles $R_n$ we use phenomenological 
quark models. The results discussed in detail below are based on a model that exhibits a 
linear confinement potential. Note that we repeated all calculations also with a quadratic
confinement potential and found that, while there are differences in the details, the pattern of the 
level spacing has no influence on the general behavior of the pole trajectories discussed in the paper.
 
To make things as simple as possible, we consider the large-$n$ limit and approximate the masses $M_n$ as
\begin{equation}
M_n=2M_Q+E_n,
\end{equation}
where $E_n$ is the $n$-th eigenenergy of the massless Hamiltonian 
\begin{equation}
H=2p_r+\sigma r.
\end{equation}
Here $p_r$ is the radial component of the momentum operator and $\sigma$ is the string tension. 
With the help of the quasiclassical quantization condition the eigenspectrum $\{E_n\}$ can be found in a simple 
analytic form (see, \emph{e.g.}, the discussion in Ref.~\cite{Kalashnikova:2001ig}),
\begin{equation}
E_n^2=4 \pi \sigma (n-1),\quad n = 1, 2, \ldots, N,
\end{equation}
where we dropped the zero-point energy such that the energy is 
counted from the ground state level with energy $E_1$. We therefore arrive at a very simple analytical formula for the mass spectrum of 
the resonances $R_n$,
\be
M_n=2M_Q+\sqrt{4\pi\sigma (n-1)},\quad n = 1, 2, \ldots, N,
\label{Mn}
\ee
which will be used in numerical calculations below. Strictly speaking, approximation (\ref{Mn}) can only be used for 
$n\gg 1$, however, the error we make using Eq.~(\ref{Mn}) for the states with $n\sim 1$ does not affect the 
phenomena under study. 

In order to proceed we need to specify the parameters of the model.
We again stick to values that are phenomenologically adequate in the charm sector, 
\begin{equation}
\sigma=0.16~\GeV^2,\quad M_Q=1.7~\GeV,\quad m=2~\GeV
\end{equation}
for the string tension $\sigma$, the heavy-quark mass $M_Q$, and the mass of the field $\varphi$. In particular, $m>M_Q$ 
complies with the 
interpretation of the field $\varphi$ as a heavy-light meson containing the heavy quark $Q$. Also, for such masses, the 
$\varphi\bar{\varphi}$ threshold lies at $2m=4$~GeV while the lowest states $R_1$ and $R_2$ have the masses 3.4~GeV and 
4.8~GeV, respectively, that 
is $R_1$ appears below the threshold while all other $R_n$'s, with $n>1$, lie above it. This way we can study the 
behavior of both the bound 
(virtual) state
and the resonances with a varying coupling strength.

For the last parameters of Lagrangian (\ref{eq:Lagrangian})---namely, for the coupling constants $g_n$---we consider two 
models. As Model A we employ
the simplest assumption on the behavior of the coupling constants $g_n$ and treat them as energy- and $n$-independent 
constants. Therefore, in model 
A we use
\be
\mbox{Model A}:\quad g_n=g
\label{modelA}
\ee
for all values of $n$. In Subsec.~\ref{sec:ExpDecreasingCouplings} we repeat our analysis for another model for the 
couplings $g_n$ and arrive at 
essentially the same conclusions. 

\subsection{Pole trajectories and the impact of renormalization}
\label{sec:Pole trajectories and the impact of renormalization}

\begin{figure*}
\begin{center}
\includegraphics[width=.7\textwidth]{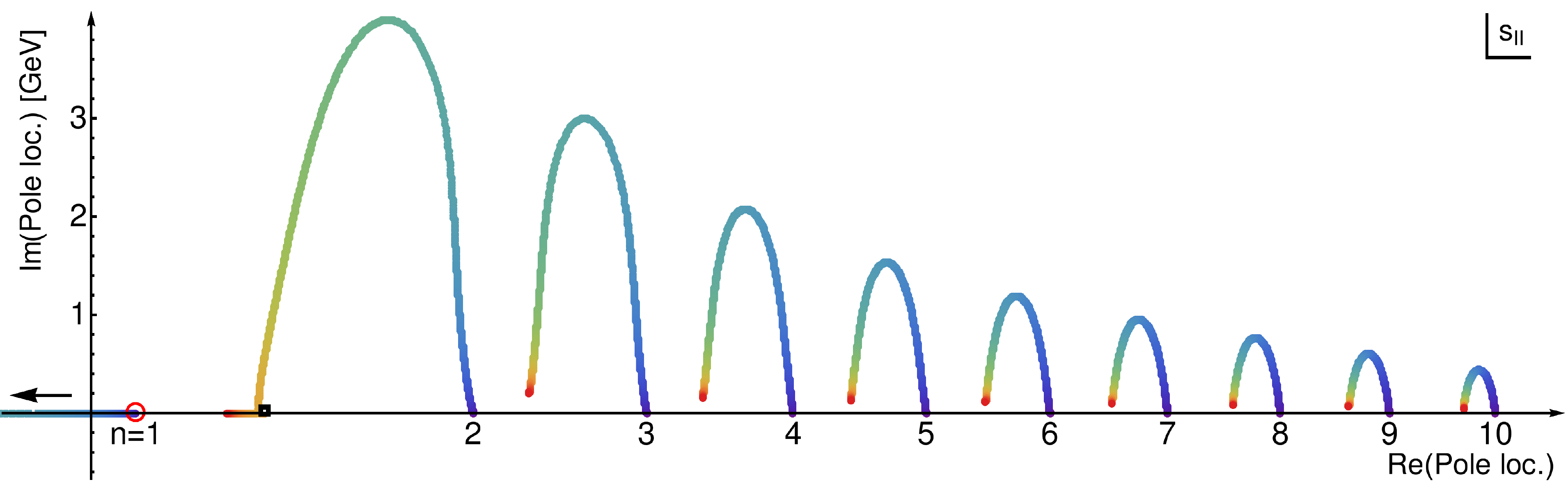}\quad
\raisebox{3mm}{\includegraphics[width=.07\textwidth]{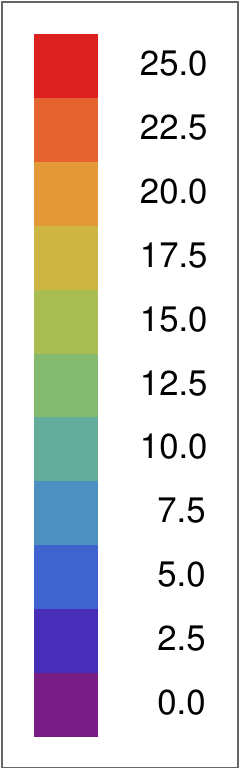}}
\end{center}
\caption{Pole trajectories in the second Riemann sheet of the complex $s$-plane for a 10-level system complying with Model A---see 
Eq.~(\ref{modelA})---for the values of $g$ 
increasing from 0 to 25~GeV. The $\Delta$-term is fixed according to Eq.~(\ref{Delta2ch}) 
($\Delta\approx0.01$). The black square indicates the position of the threshold. The red circle indicates the position of the bound state on the 
first sheet. Each trajectory is labeled by their respective principal quantum number $n$ at $g=0$.}
\label{fig:LinearPotential_N10_a0_Trajectories}
%
\begin{center}
\includegraphics[width=.7\textwidth]{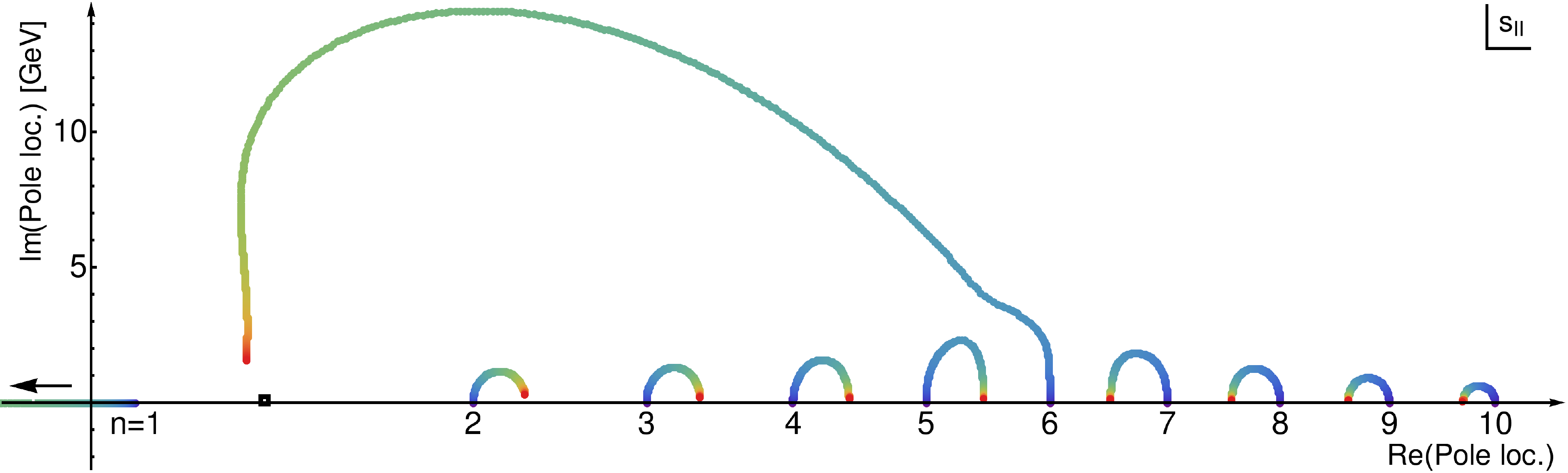} \quad
\raisebox{3mm}{\includegraphics[width=.07\textwidth]{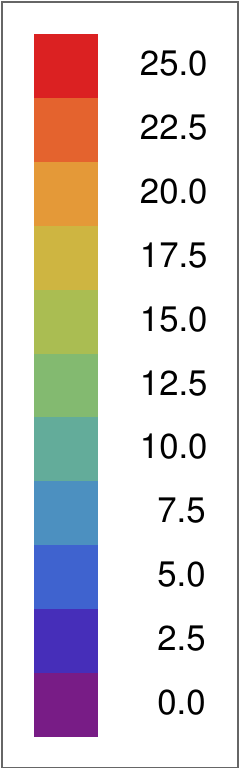}}
\end{center}
\caption{The same as in Fig.~\ref{fig:LinearPotential_N10_a0_Trajectories} but for $\Delta=0$.}
\label{fig:LinearPotential_N10_a0_D0_Trajectories}
%
\begin{center}
\includegraphics[width=.7\textwidth]{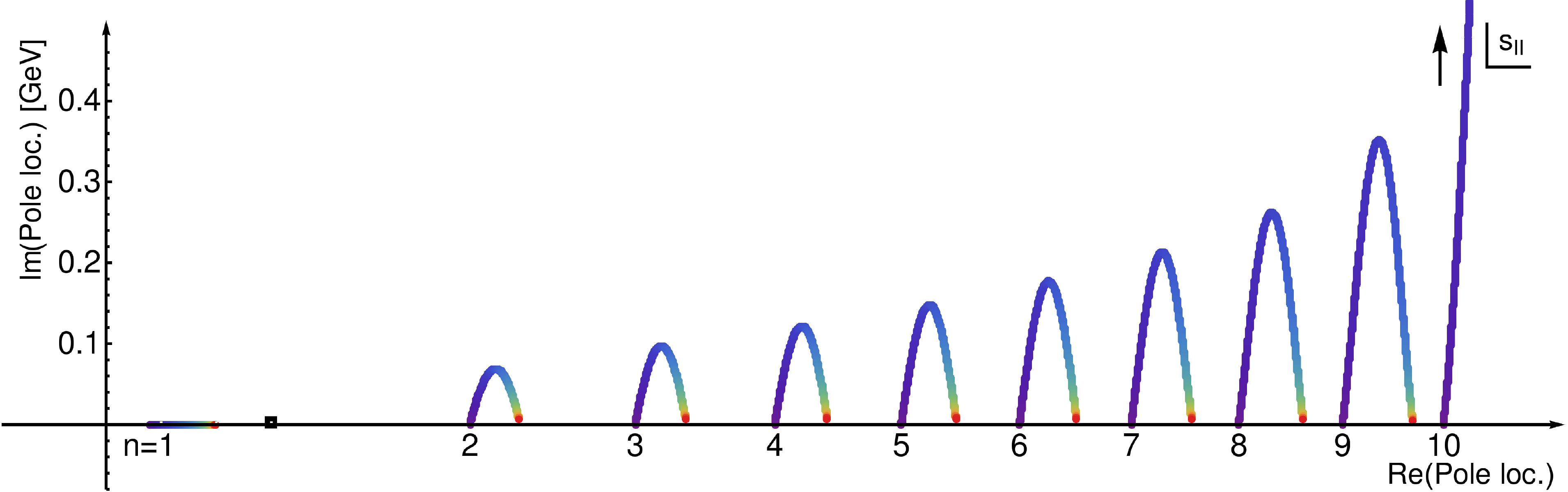}\quad
\raisebox{3mm}
{\includegraphics[width=.07\textwidth]{LinearPotential_N10_a0_D0_Legend.pdf}}
\end{center}
\caption{The same as in Fig.~\ref{fig:LinearPotential_N10_a0_Trajectories} but for $\Delta=-0.1$. }
\label{fig:NonRelativistic_VaryDelta_N10_D0_1}
\end{figure*}

It was argued and illustrated in the previous Section that the renormalization condition strongly affects the pole
trajectories. Now we come to a 
systematic study of this effect in multi-resonance systems which allows us to identify a new phenomenon related to a 
collective behavior of the 
poles. To this end we use the toy model described above and stick to Model A for the 
couplings---see Eqs.~(\ref{eq:Lagrangian}) and (\ref{modelA}), respectively. 

We start considering a multi-resonance model (\ref{eq:Lagrangian}) with $N\gg 1$---for illustration we stick to $N=10$ 
and $N=20$---and 
use three different ways to fix the $\Delta$-term in the loop operator defined in Eq.~(\ref{piofs}) (to understand the
origin of this freedom we refer the reader
to the discussion below Eq.~(\ref{piofs})):
\begin{enumerate}
\item[(i)] we keep the bound state fixed in its original position ($cf$. Eq.~(\ref{Delta2ch}));
\item[(ii)] we choose $\Delta=0$---this condition may be understood as absorbing the real part of
the loop into the masses of the resonance states;
\item[(iii)] we set $\Delta=-0.1$ to illustrate better the impact of the renormalization condition on the
pole trajectories.
\end{enumerate}
Trajectories of the poles for cases (i), (ii), and (iii) above are shown in 
Figs.~\ref{fig:LinearPotential_N10_a0_Trajectories}, 
\ref{fig:LinearPotential_N10_a0_D0_Trajectories}, and \ref{fig:NonRelativistic_VaryDelta_N10_D0_1}, respectively. One 
can draw several conclusions from 
these figures. To begin with, the poles show the same behavior as in the simple examples from the previous 
Section. For $g=0$ there are two degenerate poles below the threshold, located at $s=M_1^2$, one on the first sheet 
(bound state) and one of the second sheet (virtual 
state), and, in addition, there are pairs of poles on the second sheet at $s=M_n^2$ ($n>1$).
For the coupling $g$ departing from 0 the latter poles with $n>1$ acquire 
imaginary parts which, for small values of $g$, grow with increasing values of $g$. However, as the coupling 
exceeds some 
critical value, the pole trajectories bend and start to approach the real axis again. As a 
result, the resonances decouple from the continuum for large values of $g$.

Meanwhile, the behavior of the poles for $n=1$ deserve special attention. The
fate of the bound state pole and its virtual mirror state  depends strongly on the assumption 
made 
for the subtraction term $\Delta$. For example, in case (i) the virtual state goes far away from the threshold as $g$ grows while in case 
(iii) 
the virtual level and the bound state have nearly equal masses for all values of the coupling. 

In addition, some of the resonance poles with $n>1$ demonstrate a striking behavior---see 
Figs.~\ref{fig:LinearPotential_N10_a0_Trajectories}-\ref{fig:NonRelativistic_VaryDelta_N10_D0_1}. In case (i) all 
resonance trajectories 
look similar and the poles behave as was just explained above. However the $n=2$ trajectory hits the real axis just 
below the threshold thus 
producing two virtual states of which one proceeds toward the threshold along the real axis. In the meantime, in case 
(ii), the trajectory with $n=6$ deviates severely from the ``normal'' behavior. Indeed, this state acquires a far larger width compared to the other 
resonances before it turns around. 
Also, its real part spans a far wider range. In the last case, (iii), an unusual behavior is exhibited by the highest 
state with $n=10$. 

The result reported in this Section are obtained for the scalar loop taken to leading order in the center-of-mass 
momentum. If the full expression for the relativistic scalar loop is used instead, an additional nonanalyticity appears 
on the unphysical sheet and consequently the trajectories change quantitatively. However, qualitatively
we observe the same phenomena and thus the conclusions stay unchanged.

To better understand the physics underlying the nontrivial behavior described above, in the next Section we study the residues of the 
physical propagators of the field $R_n$.

\subsection{Residues and collective phenomena}

\begin{figure*}
\centering
{\includegraphics[width=0.3\textwidth]{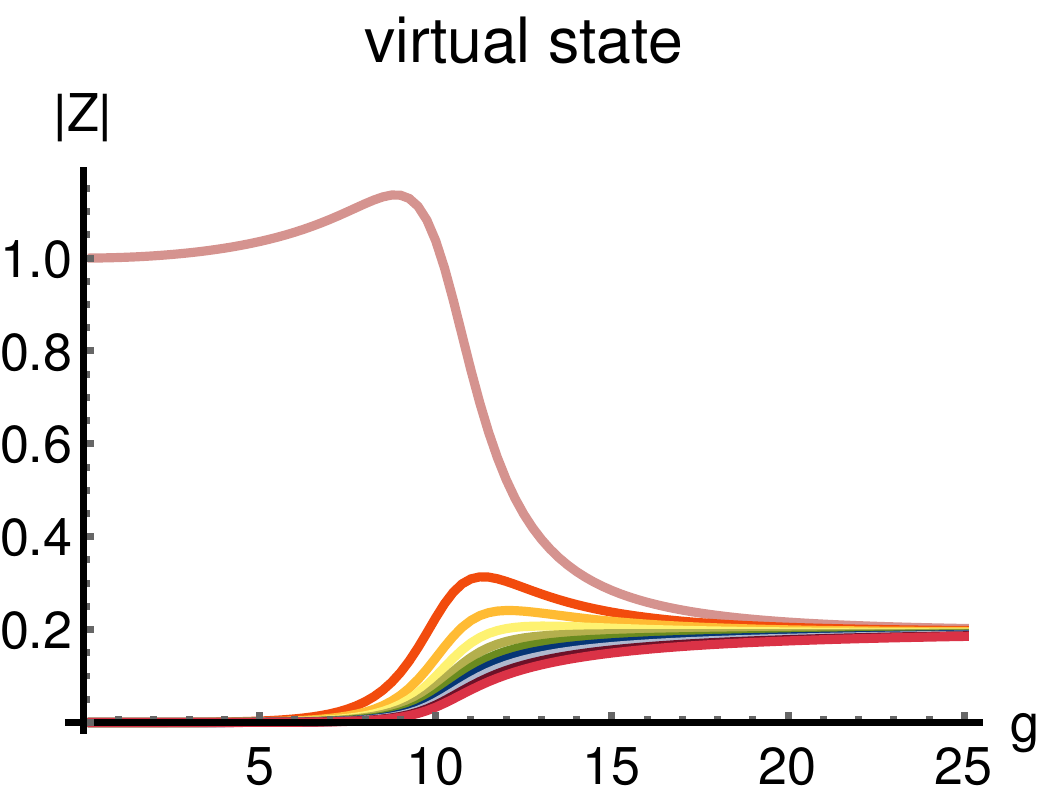}\:
\includegraphics[width=0.3\textwidth]{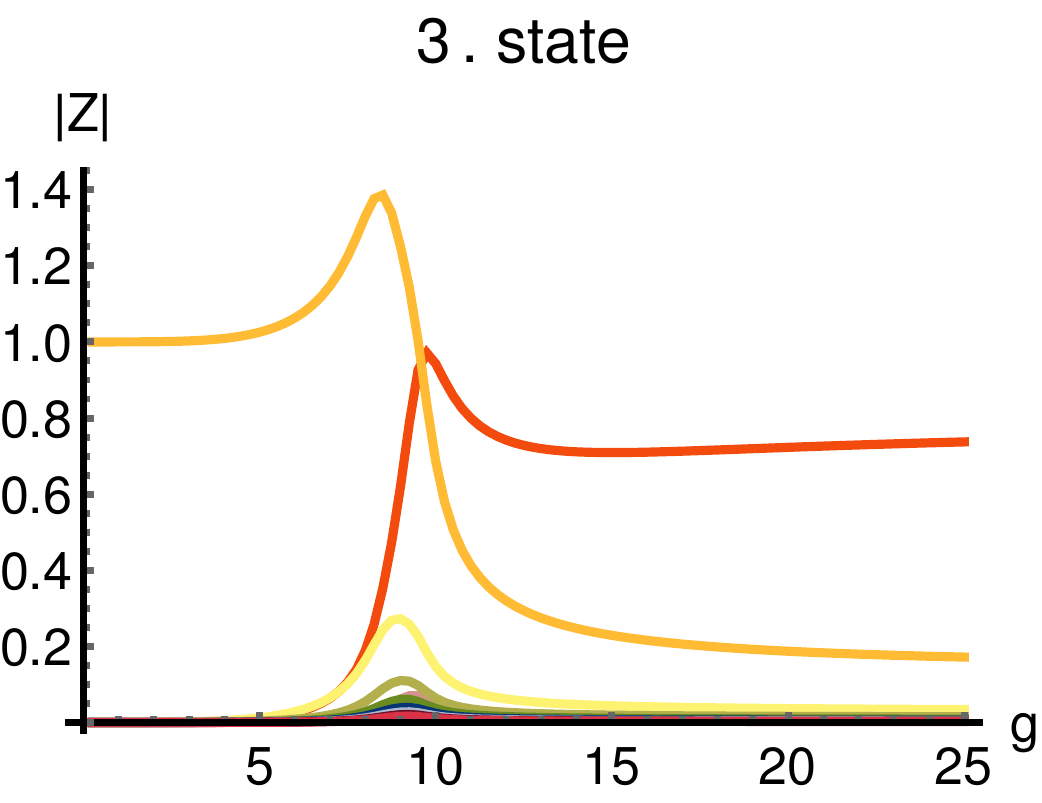}\\
\hspace{1.2cm}
\includegraphics[width=0.3\textwidth]{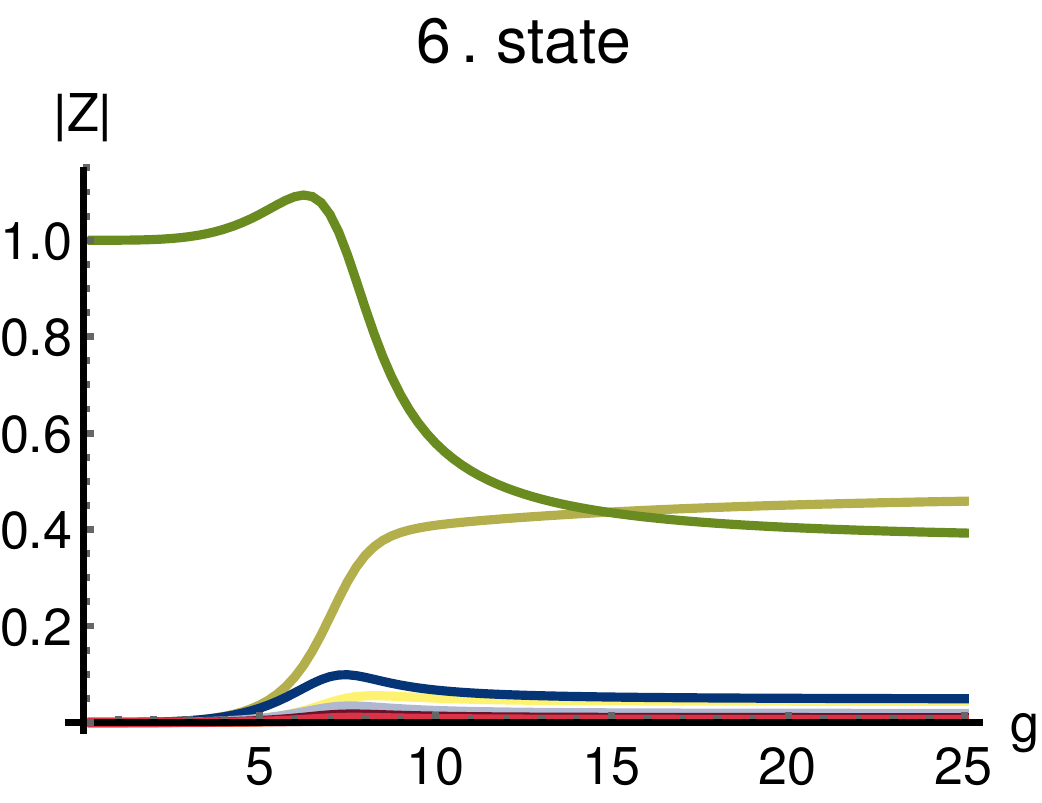}\:
\includegraphics[width=0.3\textwidth]{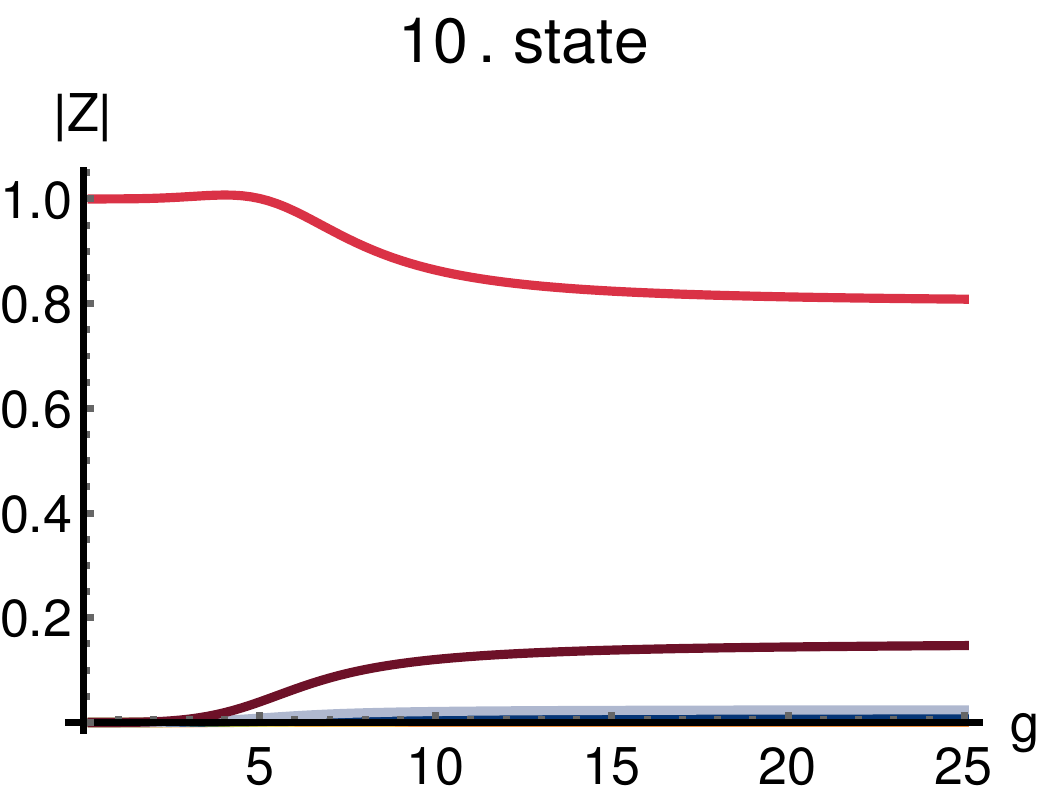}\:
\includegraphics[width=0.055\textwidth]{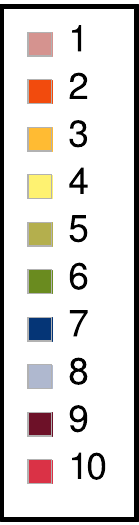}
} 
\caption{Absolute values of $Z_{nn}$ for a few resonances for the coupling $g$ growing from 0 to 25~GeV in the 
10-level 
system. The $\Delta$-term is fixed according to Eq.~(\ref{Delta2ch}) 
($\Delta\approx0.01$).}\label{fig:LinearPotential_N10_a0_Z}
\end{figure*}

\begin{figure*}
\centering
{\includegraphics[width=0.3\textwidth]{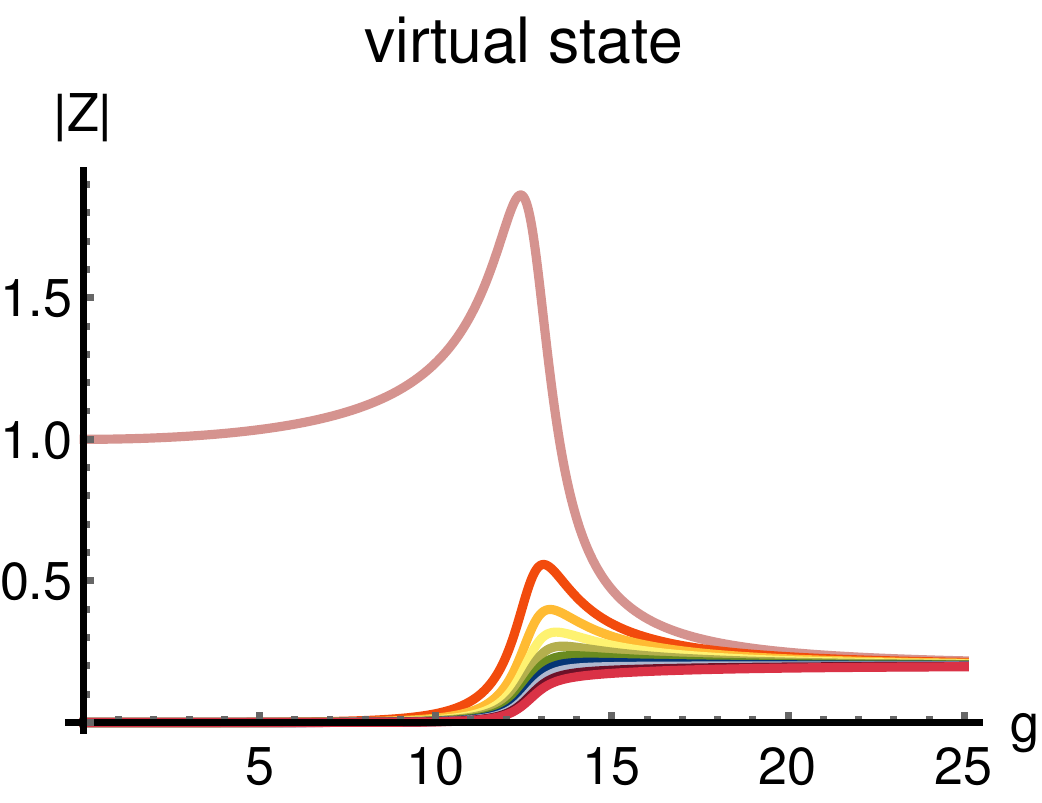}\:
\includegraphics[width=0.3\textwidth]{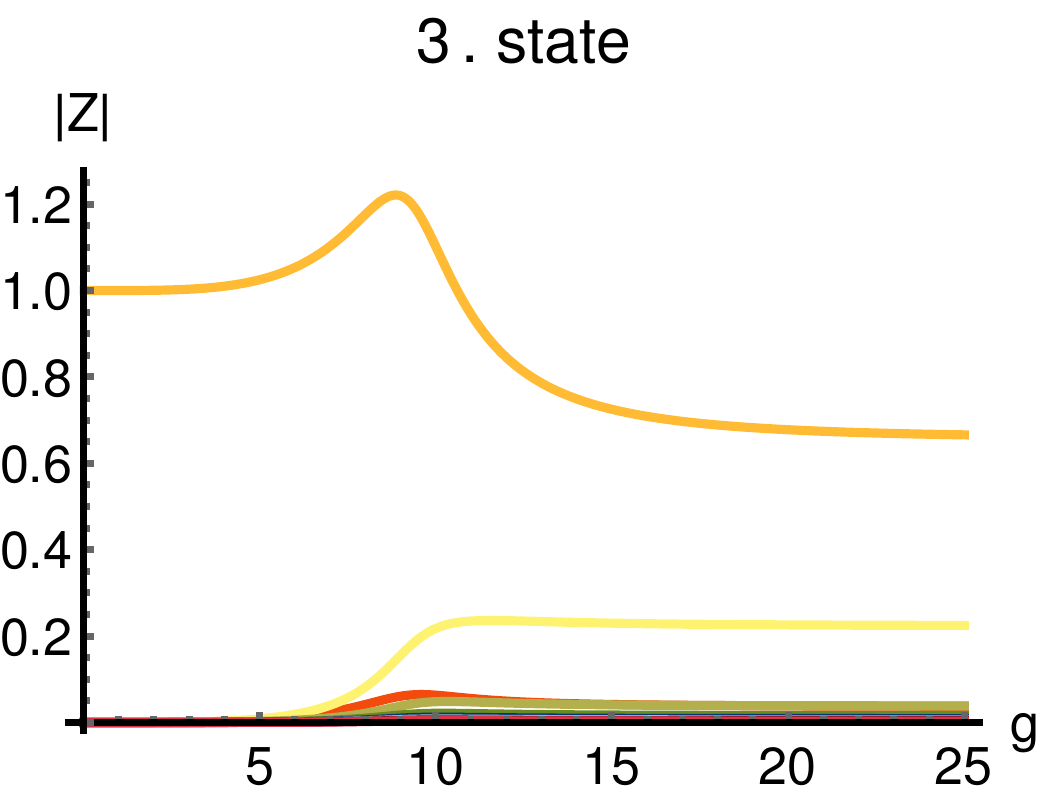}\\
\hspace{1.2cm}
\includegraphics[width=0.3\textwidth]{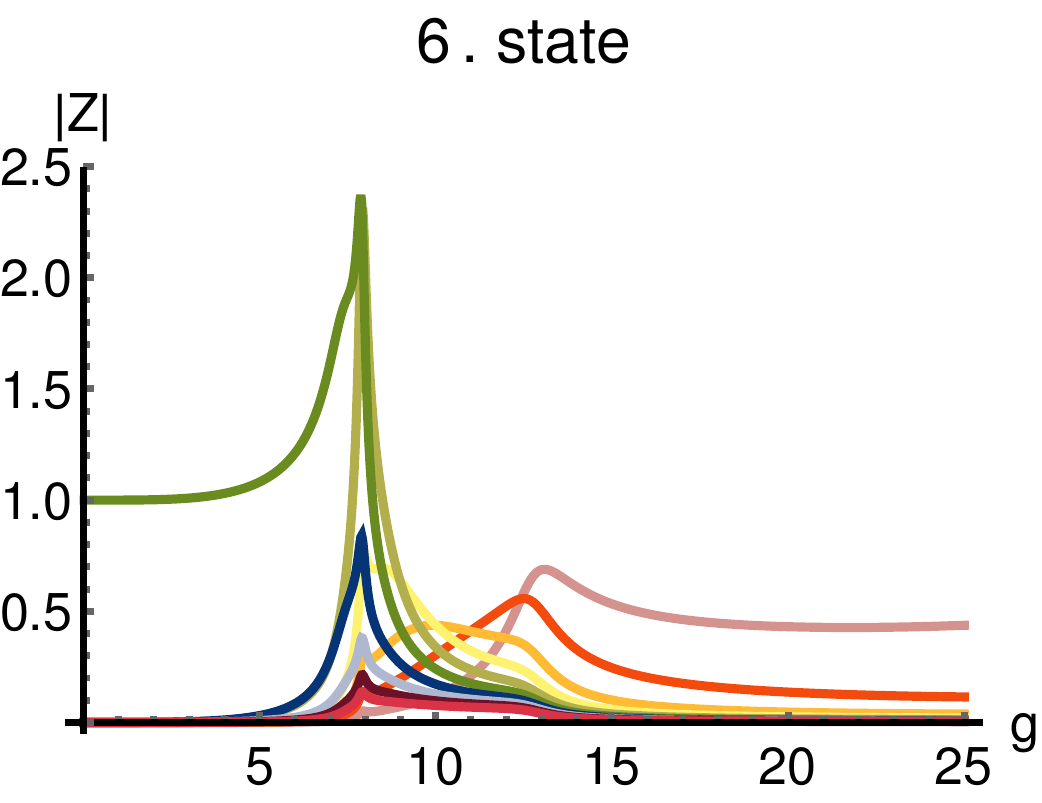}\:
\includegraphics[width=0.3\textwidth]{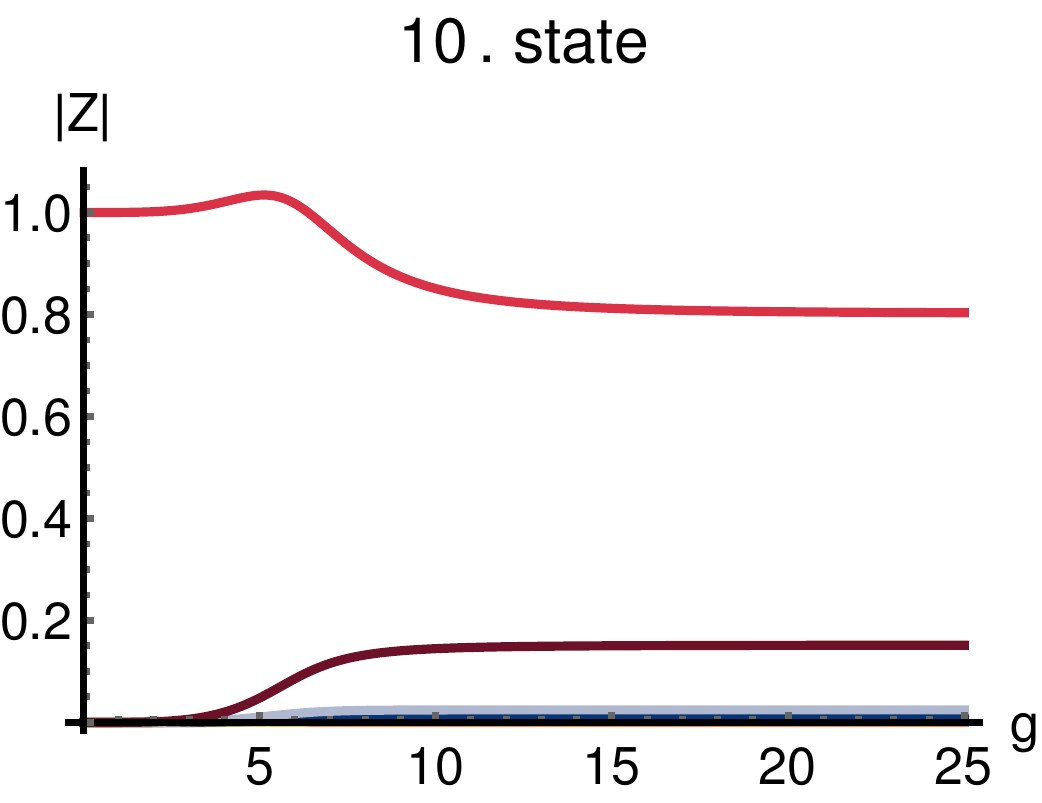}\:
\includegraphics[width=0.055\textwidth]{LinearPotential_N10_a0_Residues_Legend.pdf}}
\caption{The same as in Fig.~\ref{fig:LinearPotential_N10_a0_Z} but for $\Delta=0$.}
\label{fig:LinearPotential_N10_a0_D0_Z}
\end{figure*}

\begin{figure*}
\centering
{\includegraphics[width=0.3\textwidth]{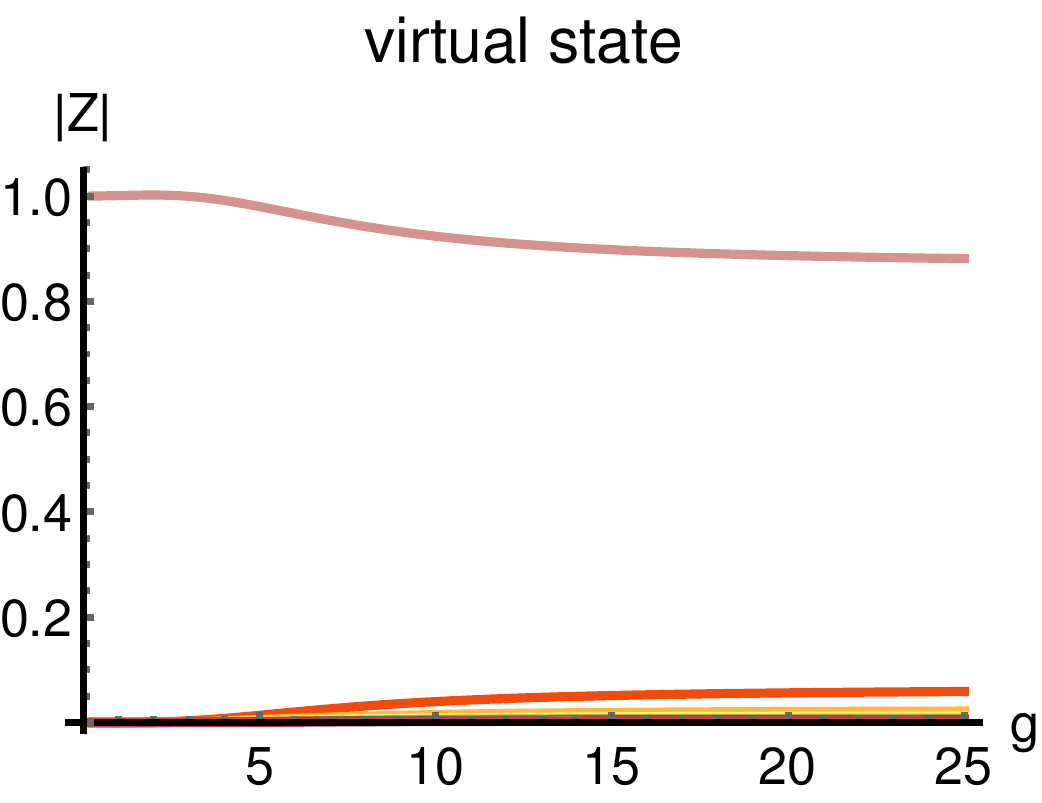}\:
\includegraphics[width=0.3\textwidth]{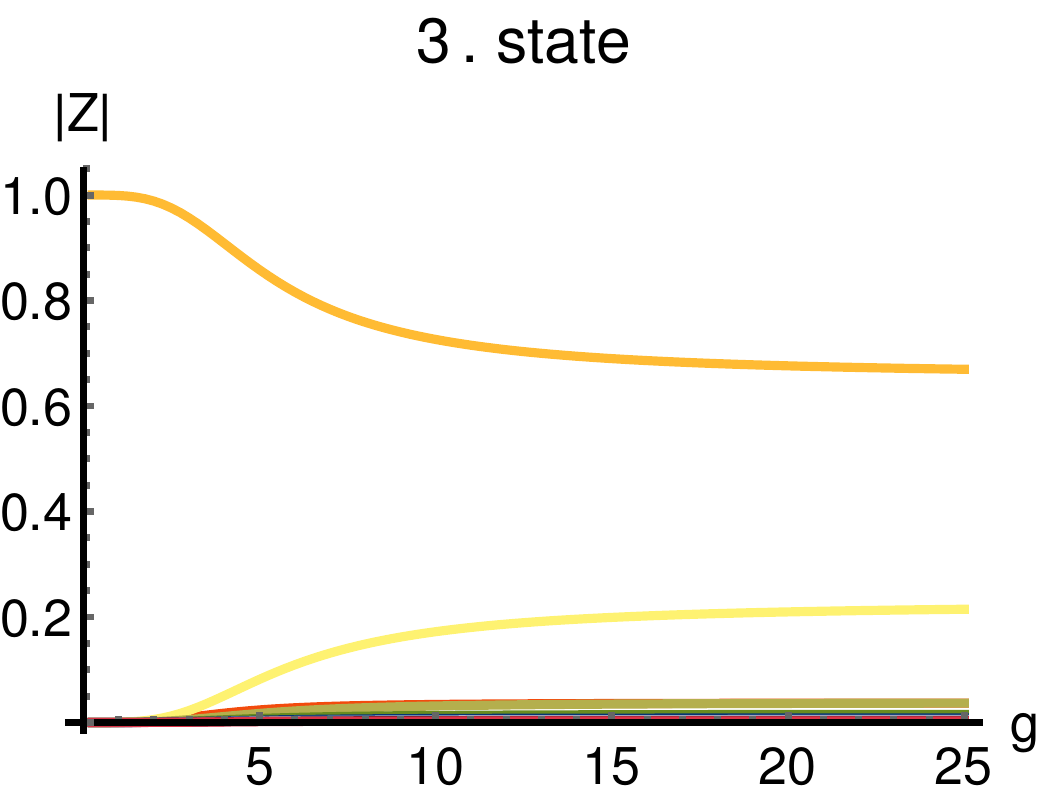}\\
\hspace{1.2cm}
\includegraphics[width=0.3\textwidth]{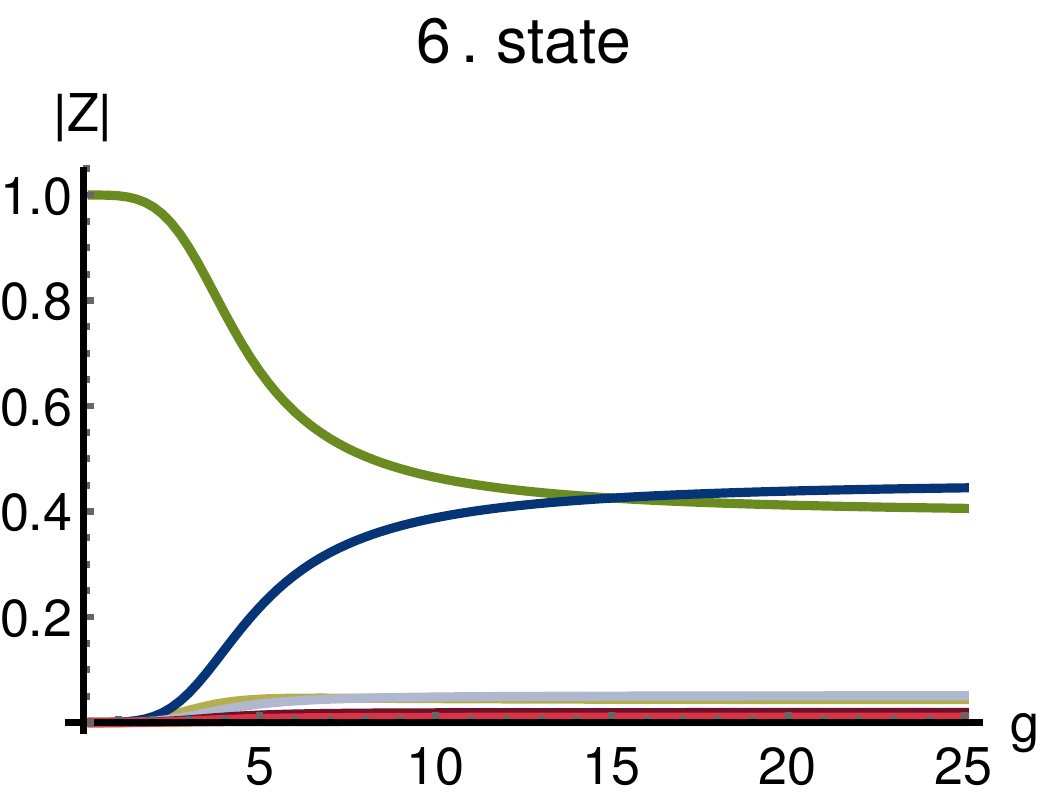}\:
\includegraphics[width=0.3\textwidth]{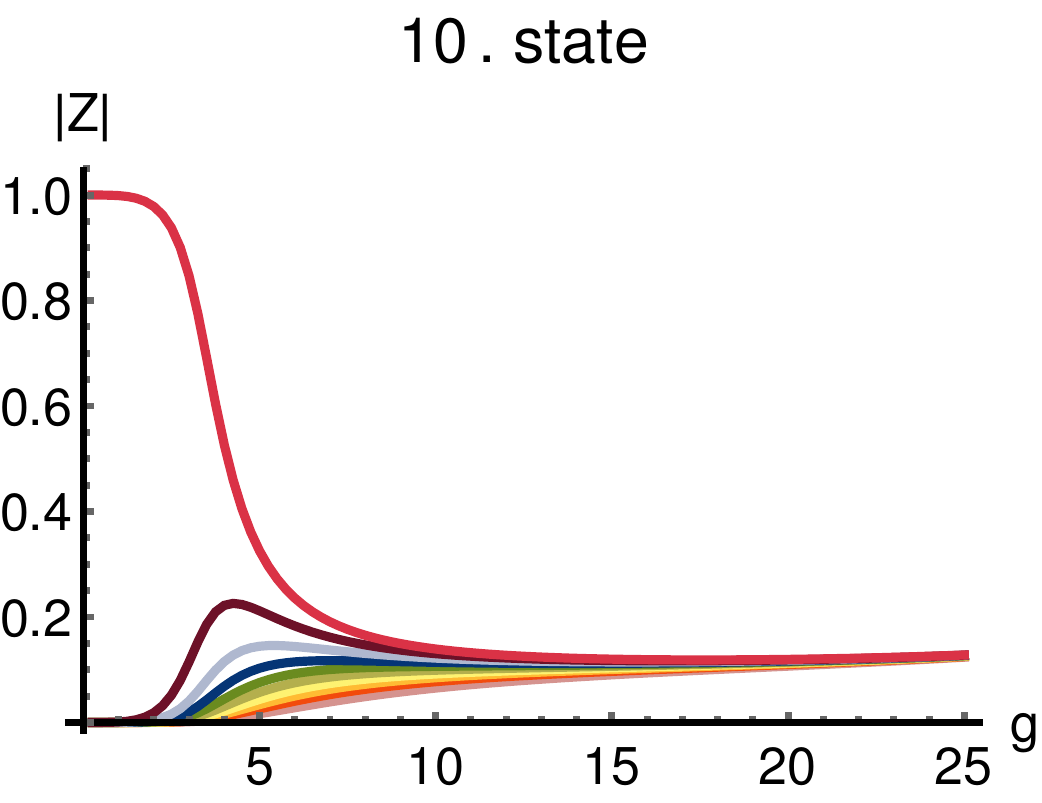}\:
\includegraphics[width=0.055\textwidth]{LinearPotential_N10_a0_Residues_Legend.pdf}}
\caption{The same as in Fig.~\ref{fig:LinearPotential_N10_a0_Z} but for $\Delta=-0.1$.}
\label{fig:NonRelativistic_VaryDelta_N10_D0_1_Z}
\end{figure*}

\begin{figure*}
\centering
{\includegraphics[width=0.28\textwidth]{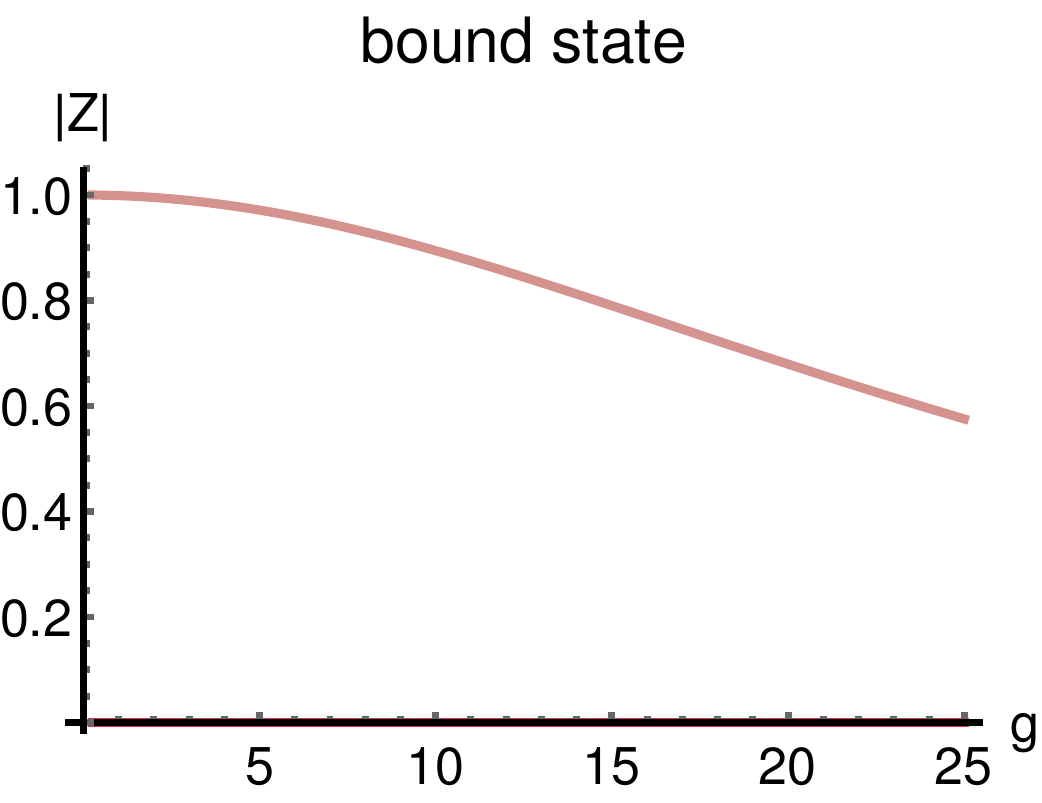}\:
\includegraphics[width=0.28\textwidth]{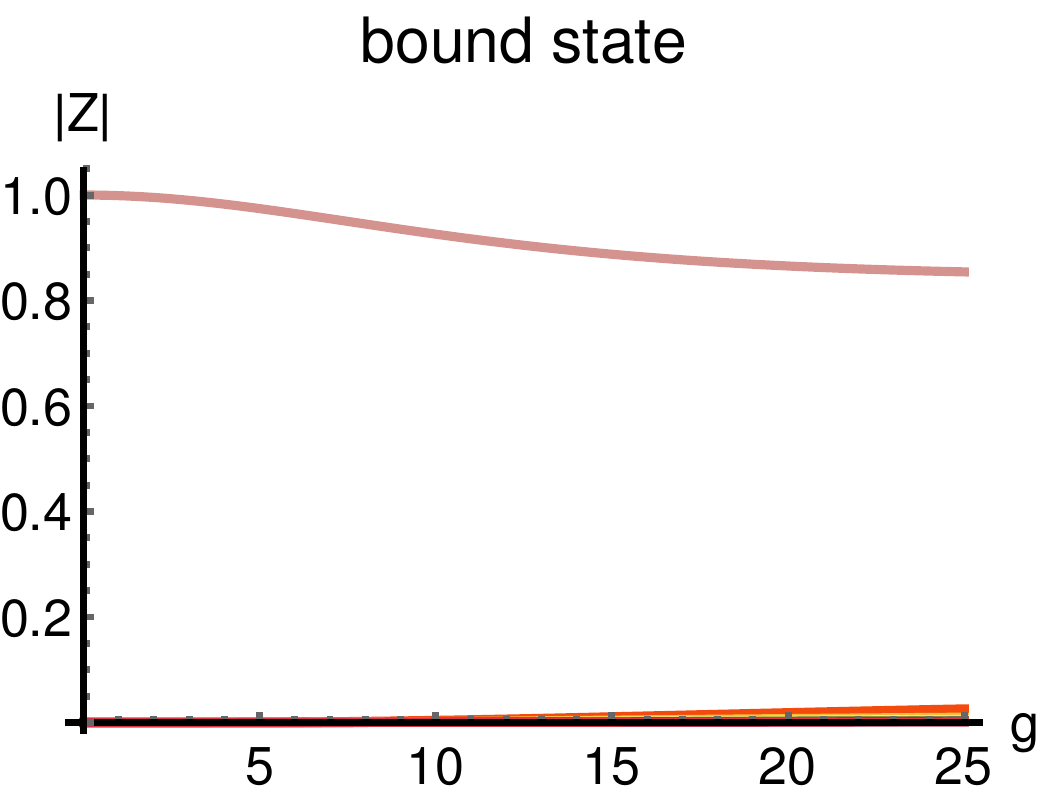}\:
\includegraphics[width=0.28\textwidth]{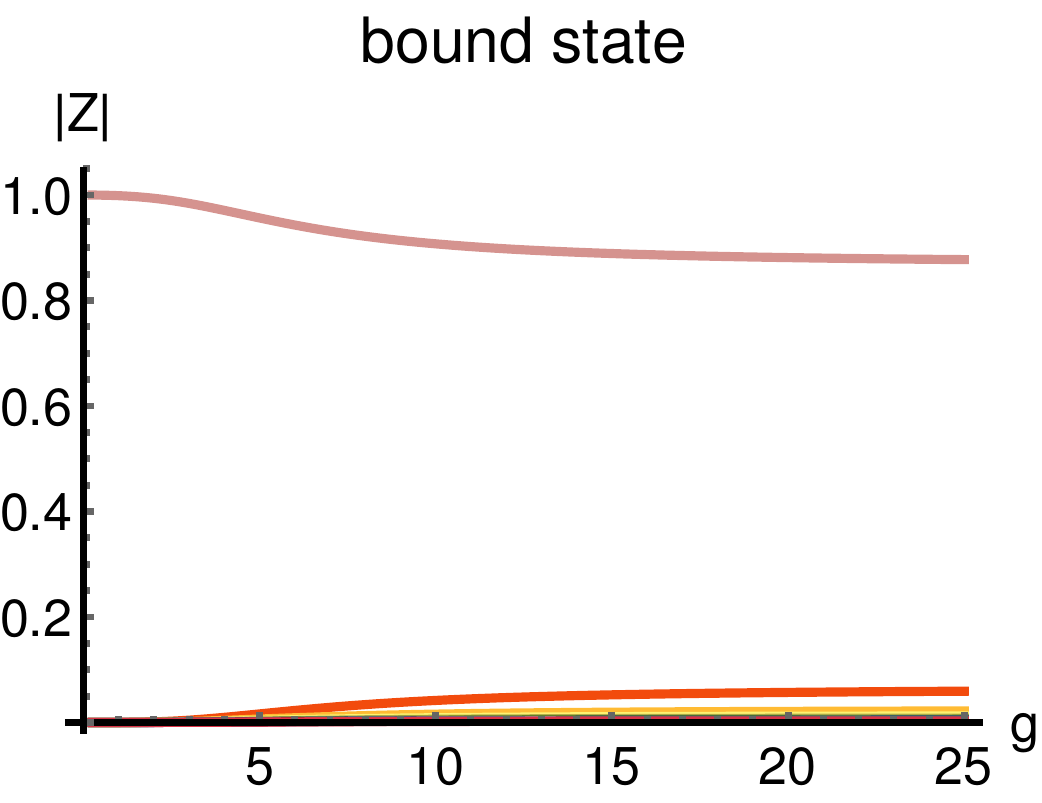}\:
\includegraphics[width=0.05\textwidth]{LinearPotential_N10_a0_Residues_Legend.pdf}}
\caption{The same as in 
Fig.~\ref{fig:LinearPotential_N10_a0_Z} ($\Delta\approx0.01$), 
Fig.~\ref{fig:LinearPotential_N10_a0_D0_Z} ($\Delta=0$) and
Fig.~\ref{fig:NonRelativistic_VaryDelta_N10_D0_1_Z} ($\Delta=-0.1$) for the bound state.}
\label{fig:LinearPotential_N10_a0_Z_BoundStates}
\end{figure*}

To better understand the physics behind the trajectories reported above, it is instructive
to investigate the residues of the $T$-matrix poles $\mathcal{R}_i$ where $s_p^{(i)}$ is the $i$-th pole 
position. These can be written in the form
\begin{equation}
\mathcal{R}_i=Res\left(T(s_p^{(i)})\right)=\sum_{n,m} g_n g_{n'} Z_{nn'}^{(i)},
\end{equation}
where the field renormalization factors $Z_{nn'}^{(i)}$ were introduced. Since we shall always focus on a particular pole, from now on we drop index 
$i$ for simplicity.

The propagators of the physical states are given by the solutions of the matrix Dyson equation
\begin{equation}
{\cal G}_{nn'}(s)=\delta_{nn'} G_n(s){+}\sum_{k}G_n(s) g_n \Pi(s) g_k {\cal G}_{kn'}(s),
\label{eq:Dyson_equation}
\end{equation}
where $G_n(s)=(s-M_n^2+i0)^{-1}$ and the explicit form of the scalar loop, $\Pi(s)$, is given to leading order in the meson momentum 
in Eq.~(\ref{piofs}). Near the pole $s_p$ one can write
\begin{equation}
{\cal G}_{nn'}(s\sim s_p)=\frac{Z_{nn'}}{s-s_p}+\text{regular terms},
\end{equation}
so that $Z_{nn'}=\lim_{s\rightarrow s_p}(s-s_p){\cal G}_{nn'}(s)$. Obviously, $s_p=M_n^2$ and $Z_{nn'}=\delta_{nn'}$ for $g=0$.
For nonvanishing couplings $g$, the solution of the Dyson equation for the $n$-th diagonal term reads
\begin{eqnarray}
{\cal G}_{nn}(s)&=&\frac{1}{s-M_n^2-g_n^2\Pi_n(s)},\nonumber\\[-2mm]
\\[-2mm]
\Pi_n(s)&=&\frac{1}{\Pi^{-1}(s)-\sum_{k\neq n} g_k^2 G_k(s)}.\nonumber
\end{eqnarray}
From this representation one can extract a simple formula for the field renormalization factors (for $s_p\neq M_n^2$),
\begin{equation}
Z_{nn'} = \frac{g_n g_{n'} G_n(s_p) G_{n'}(s_p)}{\sum_k g_k^2 G_k^2(s_p)-\Pi'(s_p)/\Pi^2(s_p)} .
\end{equation}
Alternatively those can also be determined numerically.

For a bound-state pole the $Z$ factor takes real values between 0 and 1 and, at least for very near threshold poles, 
it admits a clear physical interpretation as the probability to observe 
the bare state in the bound-state wave function \cite{Weinberg:1962hj}. 
Such a straightforward probabilistic interpretation does not exist for other poles.\footnote{Generalizations of the Weinberg approach to coupled 
channels as well
as resonances can be found in Refs.~\cite{Baru:2003qq,Hyodo:2013nka,Sekihara:2014kya,Aceti:2014ala,Guo:2015daa}.}
Nevertheless, in what follows, we look at the absolute values $|Z_{nn}|$ 
and dare to interpret them, qualitatively and with a lot of caution, as a measure of the admixture of the $n$-th bare state in the given
physical state. In particular, the values of $|Z_{nn}|$ for some representative states are shown in 
Figs.~\ref{fig:LinearPotential_N10_a0_Z}, \ref{fig:LinearPotential_N10_a0_D0_Z}, and 
\ref{fig:NonRelativistic_VaryDelta_N10_D0_1_Z} for the cases (i), (ii), and (iii), respectively.

In agreement with the observation of the previous Subsection, $Z_{nn}$ of all poles, but the 
``abnormal'' ones, behave similarly, namely, as the coupling $g$ increases, the given state starts to feel the others, however the influence 
of the remote states is much weaker than that of the closest neighbors. 

Meanwhile, the situation is very different for the unusual states, which feel not only the closest neighbors but typically all the other, more remote states. 
This kind of collectivity is seen from the behavior of the functions $|Z_{nn}|$ which take comparable values not only for the neighboring poles, but for 
the remote poles too. This is in line with the observation made in the previous Section that the trajectories of the extraordinary states 
are not localized and span a wide range. 

To complete the picture in Fig.~\ref{fig:LinearPotential_N10_a0_Z_BoundStates} we show the residues for the bound state
pole for the different values of $\Delta$ employed above. We confirmed that the bound state
wave function is properly normalized for all values of $g$. For the parameter settings used here one can see that the residues
are by far dominated by the bare pole that provided the bound state for a vanishing coupling. 
However, since the bound state poles are located far below the threshold, an application
of the Weinberg criterion to understand the compositeness of the bound state appears not possible. Meanwhile, as discussed above,
for cases (i) and (ii) the mirror state in the second sheet (virtual state) is located very asymmetrically presumably pointing at a molecular nature
of the bound state according to Ref.~\cite{Morgan:1992ge}. In contrast to this, in case (iii), the virtual state does not move and 
stays in its original location close to that of the bound state, albeit on the second sheet. Thus, in case (iii) the bound state is indeed
to be interpreted as being predominantly a quark state.

\begin{figure*}
\centering
{\includegraphics[width=.7\textwidth]{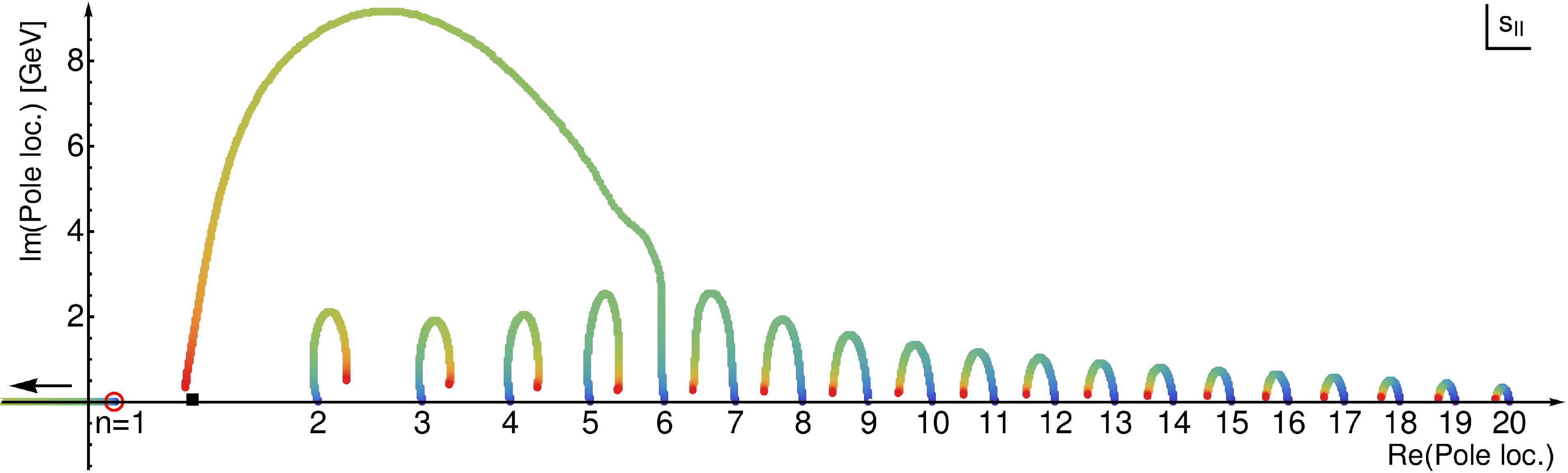}} \quad
{\includegraphics[width=.07\textwidth]{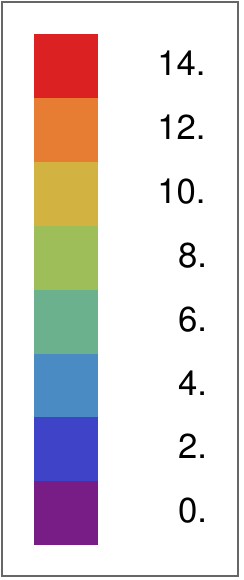}}
\caption{The same as in Fig.~\ref{fig:LinearPotential_N10_a0_Trajectories} but for $N=20$.}
\label{fig:LinearPotential_N20_a0_Trajectories}
\end{figure*}

\begin{figure*}
\centering
{\includegraphics[width=0.3\textwidth]{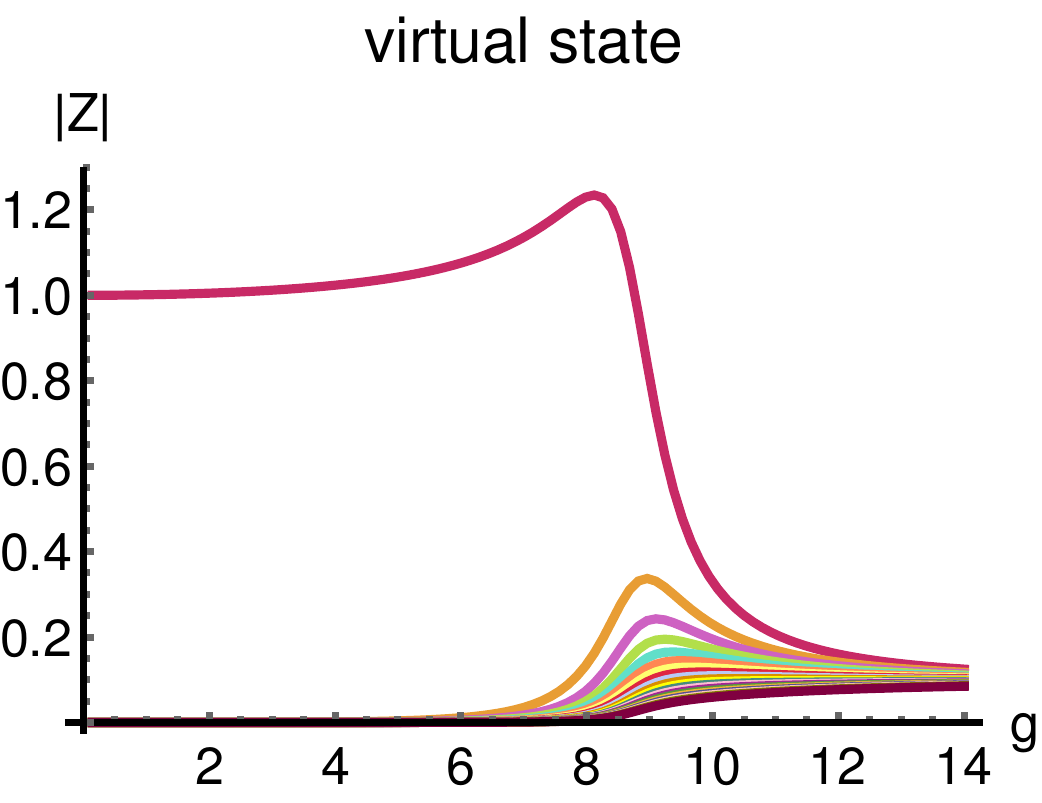}\:
\includegraphics[width=0.3\textwidth]{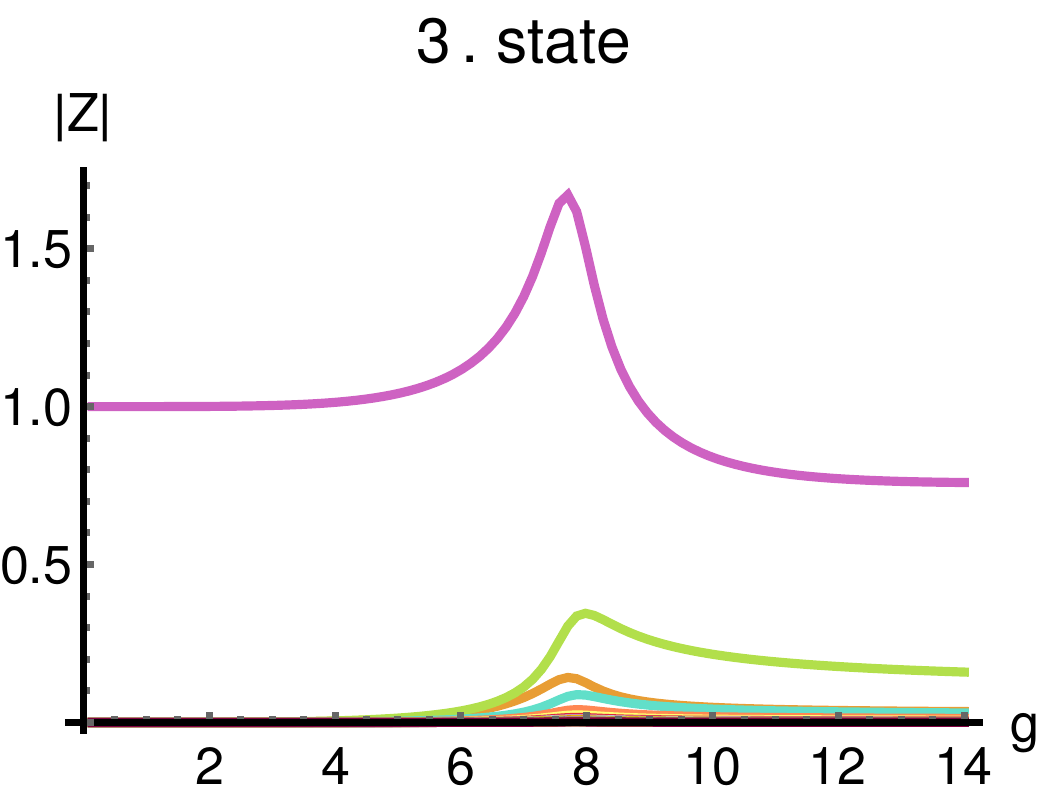}\\
\hspace{1.8cm}
\includegraphics[width=0.3\textwidth]{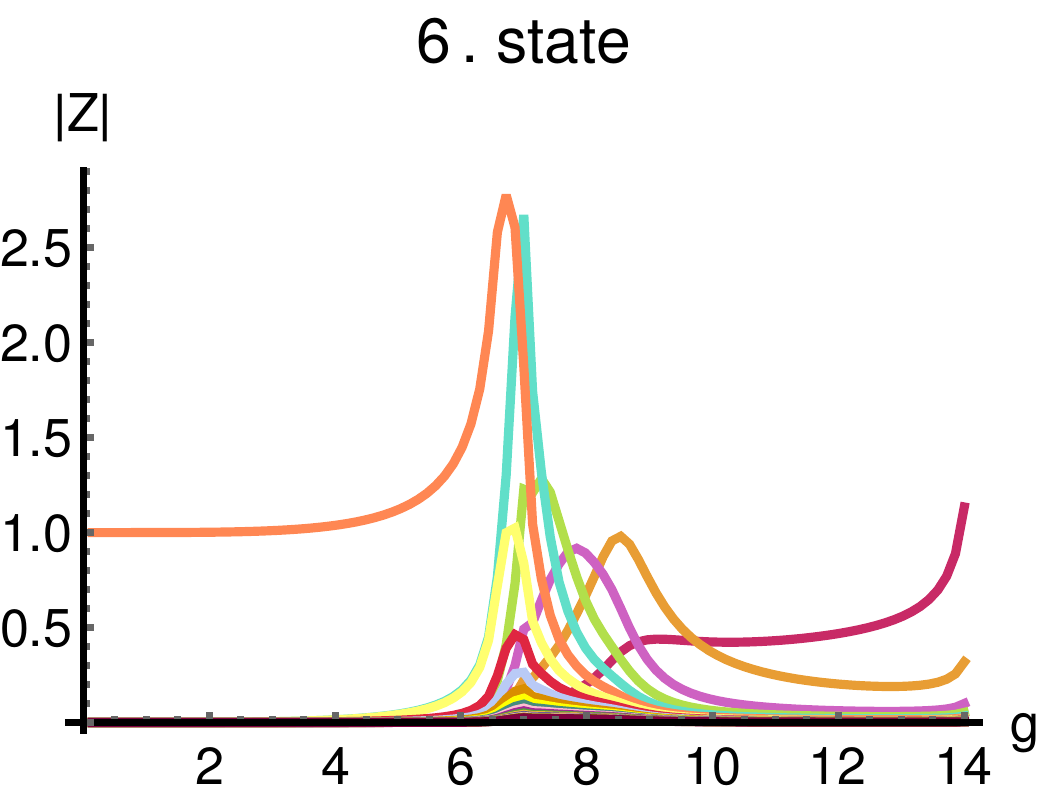}\:
\includegraphics[width=0.3\textwidth]{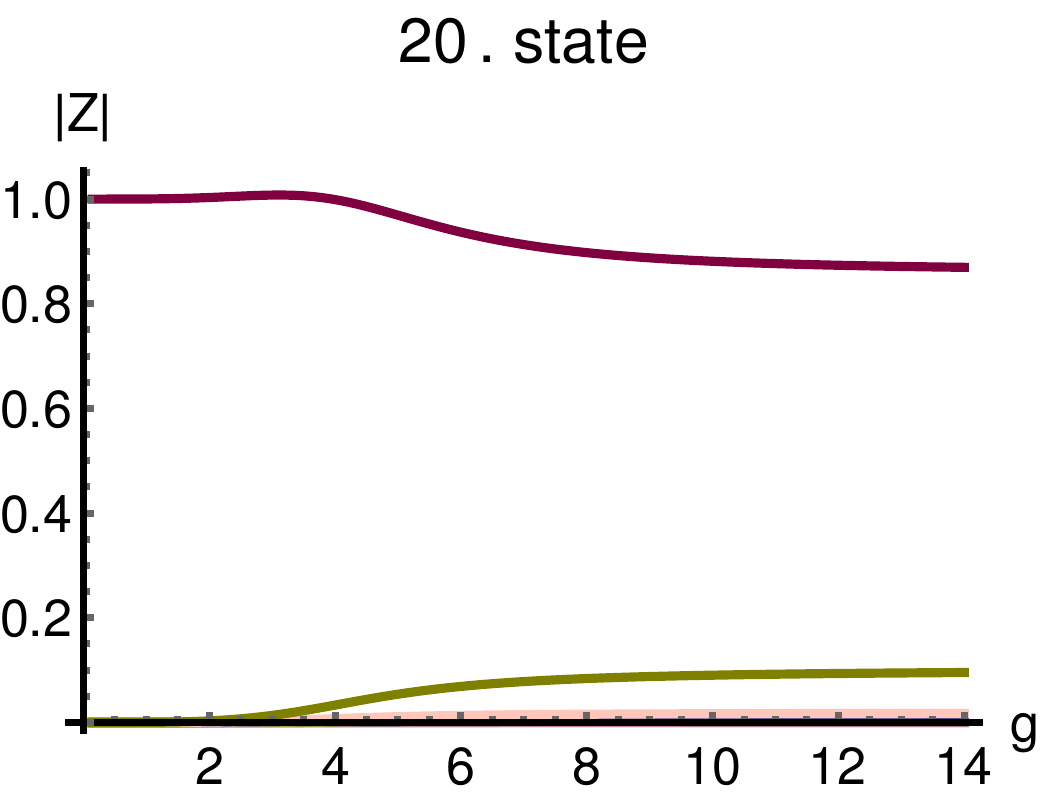}\:
\includegraphics[width=0.09\textwidth]{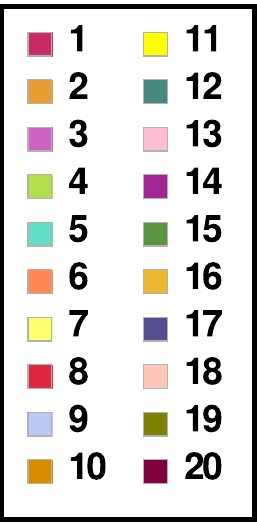}
} 
\caption{The same as in Fig.~\ref{fig:LinearPotential_N10_a0_Z} but for $N=20$. The rise of the 
$Z_nn$ at $g\to 14$ for the $n=6$ state is due to a peak which reveals itself when the trajectory hits the real 
axis. Such effects were also observed before---see, for example, 
Ref.~\cite{Nebreda:2010wv}.}
\label{fig:LinearPotential_N20_a0_Z}
\end{figure*}

To ensure that the observed collective phenomenon is not an artifact of a particular chosen number of poles $N$ we 
repeat the study outlined above for $N=20$. For illustration we show the results for the renormalization condition which keeps the bound state mass 
fixed (case (i)) in Figs.~\ref{fig:LinearPotential_N20_a0_Trajectories} and 
\ref{fig:LinearPotential_N20_a0_Z}. Clearly the system exhibits exactly the same
behavior as discussed before for $N=10$, namely that some states show a
collective behavior for $g\to\infty$ while most of the states tend to decouple from the continuum in this limit.

\subsection{Alternative model for the couplings}
\label{sec:ExpDecreasingCouplings}

\begin{figure}[t]
\begin{center}
\epsfig{file=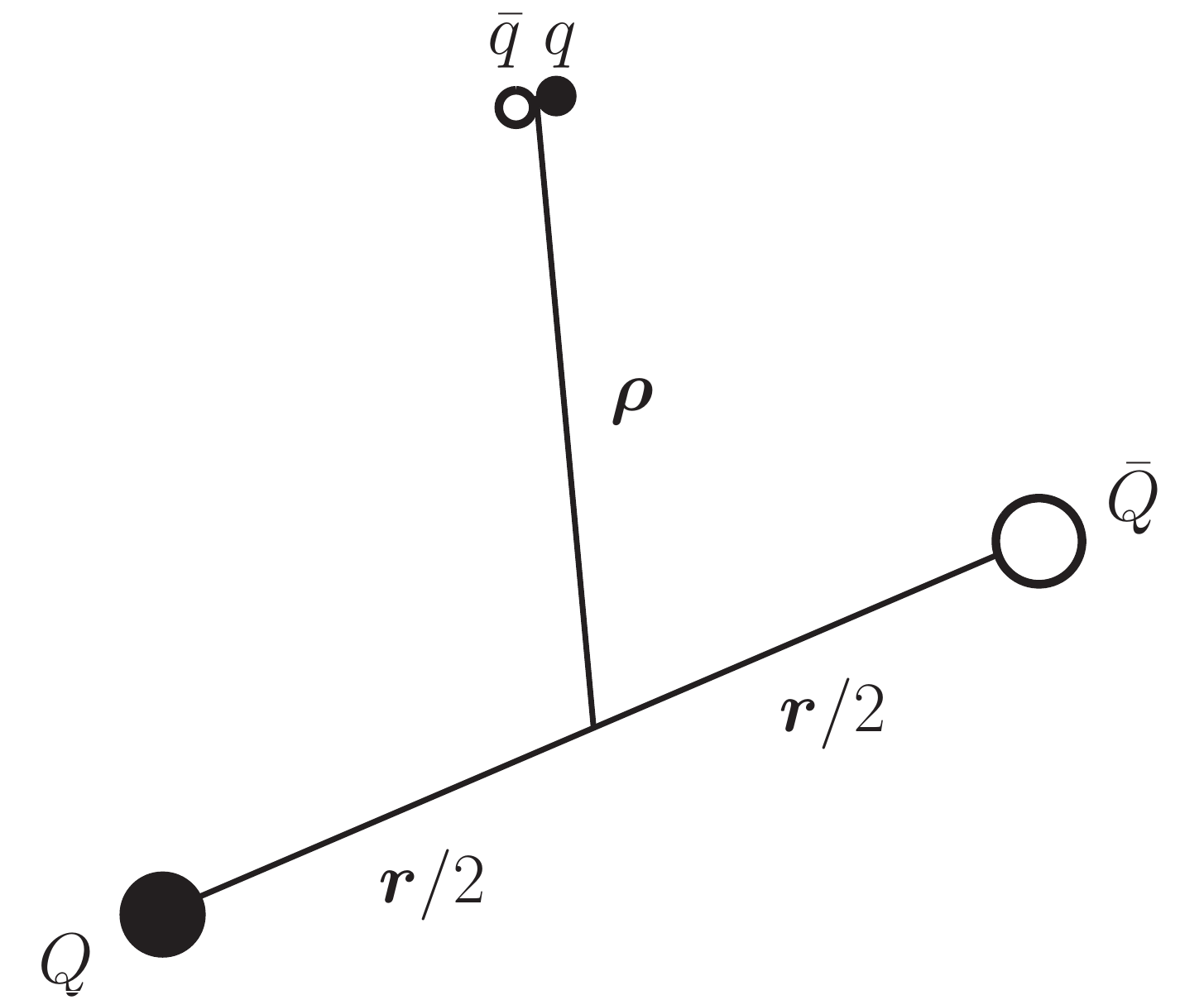,width=0.3\textwidth}
\caption{Graphical representation of the pair creation mechanism.}\label{fig:3P0}
\end{center}
\end{figure}

As the last exercise, in order to demonstrate that our qualitative findings do not depend
on the particular model chosen for the couplings, we also use a more 
realistic model for the vertex functions which we refer to as Model B. To this 
end we consider the decay of the form $R_n\to\varphi\bar{\varphi}$ and notice that, in order to model the behavior of 
the coupling $g_n$, it is sufficient 
to evaluate the overlap of the wave functions (w.f.s) in the amplitude for the light-quark pair creation through the ${}^3P_0$ 
mechanism---for 
the details see, for example, 
Refs.~\cite{Eichten:1978tg,Geiger:1994kr,Ackleh:1996yt,Liu:2011yp,Ferretti:2012zz,Ferretti:2013vua,Lu:2016mbb} and references therein,
\begin{eqnarray}\nonumber
g_n&\propto& \int d^3r 
d^3\rho\;\psi_{R_n}(\ver)\psi_{\varphi\bar{\varphi}}^*(-\ver/2)\\
& &\qquad \quad \times  \ \psi_{\varphi}^*(\ver/2+\vey)\psi_{\bar{\varphi}}
^*(\ver/2-\vey),
\label{overlap}
\end{eqnarray}
where the standard relative Jacobi coordinates $\ver$ and $\vey$ are introduced as shown in Fig.~\ref{fig:3P0}, 
$\psi_{R_n}$, $\psi_\varphi$, and 
$\psi_{\bar{\varphi}}$ are the 
bound-state w.f.s of all mesons involved, $\psi_{\varphi\bar{\varphi}}(\ver)=e^{i\vek\ver}$ is the w.f. of the free 
motion of the mesons 
$\varphi$ and $\bar{\varphi}$ with the relative momentum $\vek$, that is $\veP_\varphi=-\veP_{\bar{\varphi}}=\vek$. 

The w.f. of the heavy-heavy quarkonium $R_n$ is localized at small interquark separations while the typical size of the 
heavy-light mesons $\varphi$ 
and $\bar{\varphi}$ is much larger. Consequently the mass parameter $\beta_R$ which governs the falloff of the w.f. 
$\psi_{R_n}$ with the quark 
separation is large compared to the corresponding parameter for the meson $\varphi$, $\beta_R\gg\beta_\varphi$. As a result 
the w.f. $\psi_{R_n}(\ver)$ 
cuts the integral in $r$ at $r\ll\rho$, so that, approximately,
\begin{eqnarray}\nonumber
g_n&\propto& \left(\int d^3r e^{-i\vek\ver/2}\psi_{R_n}(\ver)\right)\left(\int d^3\rho 
\psi_\varphi^*(\vey)\psi_{\bar{\varphi}}^*(-\vey)\right)
\\
& & \quad =\int 
d^3r e^{-i\vek\ver/2}\psi_{R_n}(\ver)=\Psi_{R_n}(\vek/2),
\label{overlap2}
\end{eqnarray}
where $\Psi_{R_n}$ is the Fourier transform of the coordinate w.f. $\psi_{R_n}(\ver)$, and it was used that 
$\psi_{\bar{\varphi}}^*(-\vey)=\psi_{\varphi}(\vey)$ and that the w.f. $\psi_{\varphi}(\vey)$ is normalized. 

If, for simplicity, we stick to the quasiclassical w.f. $\Psi_R$ then it is easy to find that
\begin{equation}
g_n \propto \Psi_{R_n}(\vek/2) \propto \exp\left(-\frac{k^2}{4\sigma}\right).
\label{g_lambda_eq_4sigma}
\end{equation}
In the decay $R_n\to\varphi\bar{\varphi}$ the momentum $k$ is fully fixed by the masses of the particles in the initial 
and in the final states. 
On the other hand, in the scattering process $\varphi\bar{\varphi}\to R_n\to\varphi\bar{\varphi}$, $k$ 
is related to the invariant energy $s$ as given by Eq.~(\ref{ks}). Thus, as Model B, we use the form
\be
\mbox{Model B}:\quad g_n=ge^{-\lambda^2k^2}
\label{glambda}
\ee
which smoothly interpolates between Model A ($\lambda=0$) and formula (\ref{g_lambda_eq_4sigma}) with 
$\lambda=1/(2\sqrt{\sigma})=1.25$~GeV$^{-1}$. Strictly speaking the approximations used to 
derive Eq.~(\ref{glambda}) are valid for low-lying 
resonances $R_n$, while for $n\gg 1$ the high-order pre-exponential polynomial in the $R_n$'s w.f. starts to compete with the exponential factor so 
that $\Psi_{R_n}$ is not short-ranged any more. As a consequence, the overlap of the w.f.s in Eq.~(\ref{overlap2}) acquires a dependence on the 
excitation number $n$ and so does the coupling $g_n$ from Eq.~(\ref{glambda}). However, inclusion of this effect goes
well  beyond the scope of the present 
work, which is aimed at a qualitative investigation of the unitarization effects on the spectrum of quark states. 
We therefore leave a more refined 
treatment of the couplings $g_n$ for future publications and stick here to the simplest form
presented in Eq.~(\ref{glambda}).

The scalar loop for Model B is given by the expression
\begin{eqnarray} \nonumber
\Pi(s)&=&\frac{1}{16\pi^2m}\int_0^\infty\frac{p^2}{k^2-p^2}e^{-\lambda^2 p^2}dp\\ \nonumber
&=&-\frac{i}{16\pi m}ke^{-\lambda^2 k^2} \\
& & \qquad +\frac{1}{16\pi^2m}\mathcal{P}\int_0^\infty \frac{p^2}{k^2-p^2}e^{-\lambda^2 p^2} d p,
\end{eqnarray}
where $\mathcal{P}$ stands for the principal value prescription. As one can see, different values of $\lambda$,
referring to different quark models,
provide different values of the real part of the loop and, when parametrized in the form of Eq.~(\ref{piofs}), correspond to
different values of $\Delta$.

\begin{figure*}
\centering
\subfloat[$\lambda = 0.1$~GeV$^{-1}$]{\includegraphics[width=.75\textwidth]{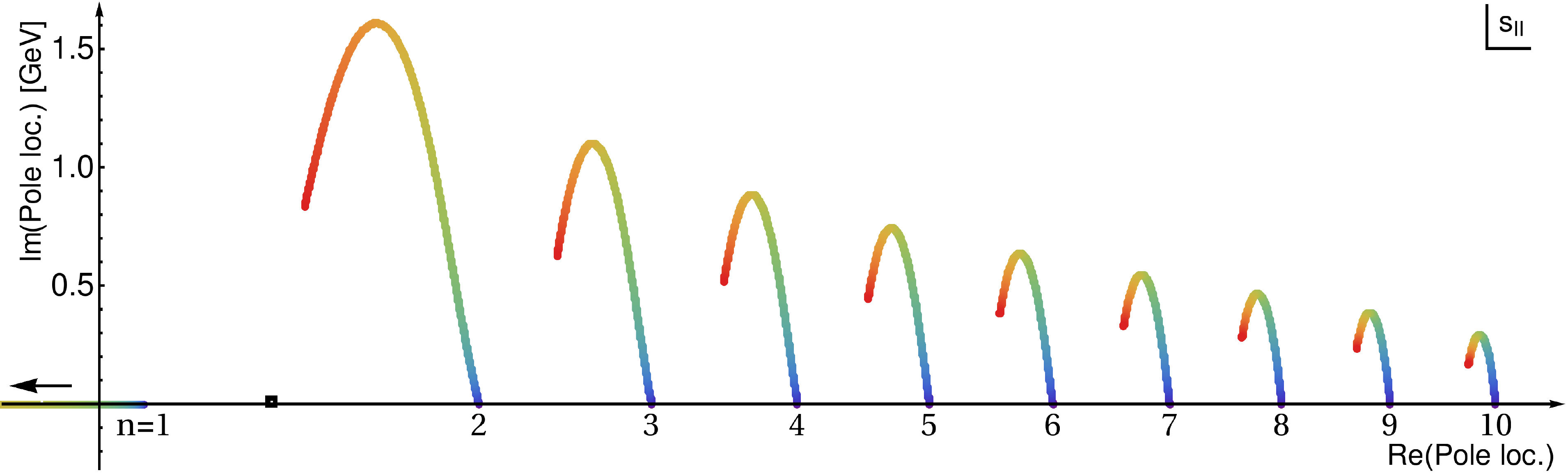}} \quad
{\includegraphics[width=.06\textwidth]{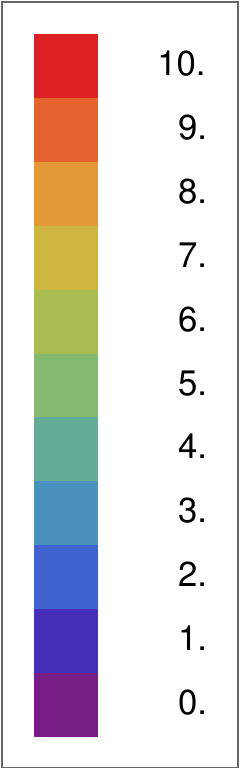}}\\
\subfloat[$\lambda = 0.3$~GeV$^{-1}$]{\includegraphics[width=.75\textwidth]{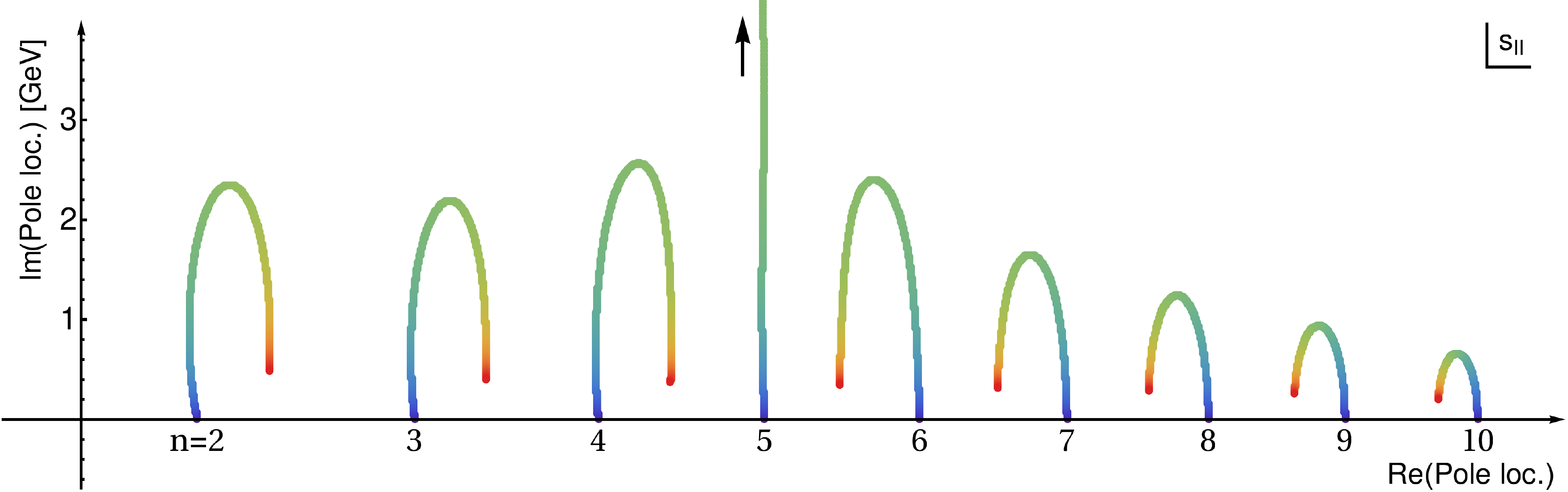}} \quad
{\includegraphics[width=.06\textwidth]{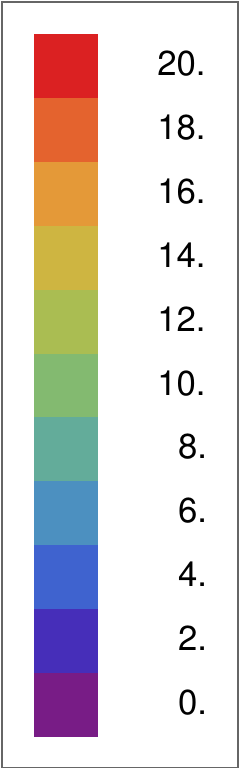}}\\
\subfloat[$\lambda = 0.7$~GeV$^{-1}$]{\includegraphics[width=.75\textwidth]{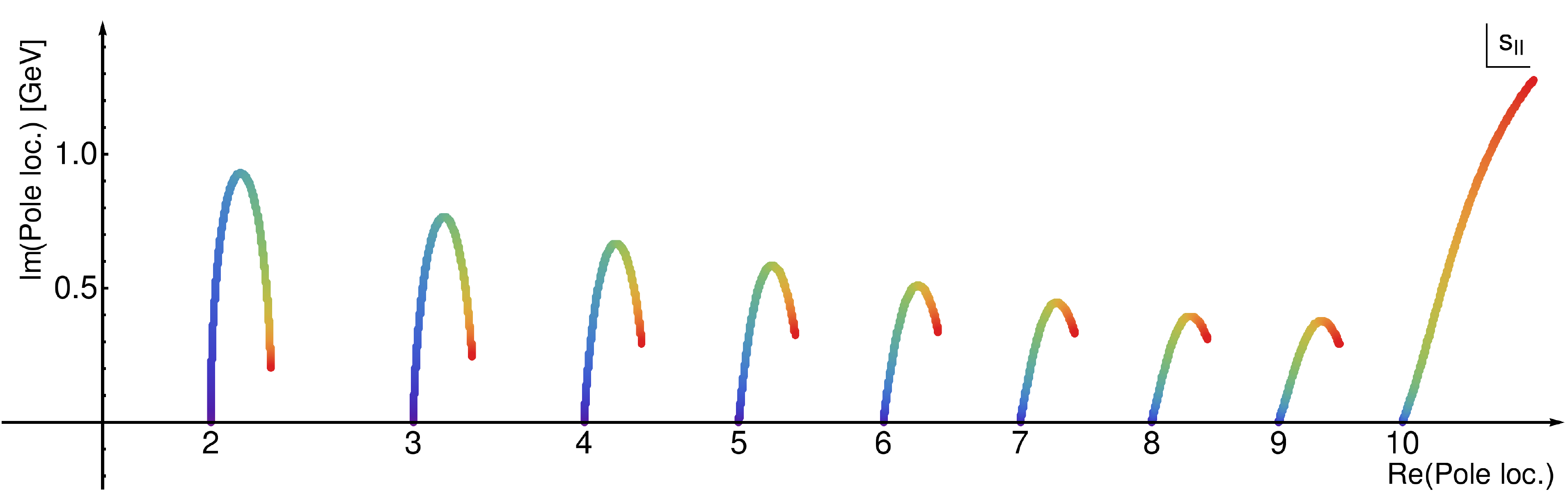}} \quad
{\includegraphics[width=.06\textwidth]{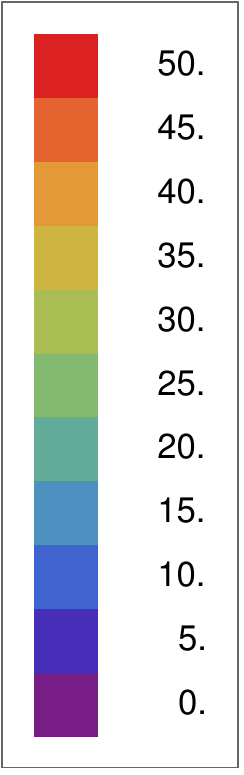}} \\ 
\caption{Pole trajectories in the complex $s$-plane for Model B with different values of $\lambda$. The black square 
indicates the threshold.}
\label{fig:LinearPot_ExpCoupling_lVary_Trajectories}
\end{figure*}

In Fig.~\ref{fig:LinearPot_ExpCoupling_lVary_Trajectories} we plot the pole trajectories for Model B with $N=10$ and with
$\lambda=0.1$~GeV$^{-1}$, $0.3$~GeV$^{-1}$, and $0.7$~GeV$^{-1}$, respectively. From these figures one can see that Model B demonstrates essentially 
the same pattern as Model A. We refrain from considering larger values of $\lambda$ as spurious poles start to appear 
in this case on both the unphysical and sometimes even on the physical sheet. For large values of the coupling $g$ 
these poles tend to approach the physical region and to interfere with the physical poles. This would make an 
interpretation of the latter questionable. Such spurious poles are artifacts of the particular parametrization used 
for the vertex function. Building a more reliable parametrization should solve the problem, however, this 
goes beyond the scope of the present paper.

\section{Interpretation}
\label{sec:int}

In this paper we found the following universal features which emerge when quark states get coupled to the 
continuum in a way consistent with unitarity, once the coupling gets too large to be treated perturbatively:
\begin{itemize}
\item[$\bullet$] As soon as the
selfenergies get of the order of the level spacing, for most states the pole trajectories bend and the width decreases
again. These states are mostly influenced by their nearest neighbors.
\item[$\bullet$] Besides the states just described at least one state possesses a pole trajectory which spans a wide (compared to the level spacing) 
range therefore acquiring contributions from multiple bare poles---we therefore call it a collective state.
\item[$\bullet$] Which state shows this collective phenomenon and if this collectivity shows up in one or more states depends
on the renormalization condition imposed and on the details of the model.
\end{itemize}
Clearly, to study the evolution of a collective pole as 
a function of some strength parameter, it is not sufficient to start from a reduced basis since in this case one omits 
states that are of relevance for the collective one. 
This does not mean, however, that it is not possible to build an effective theory with only a few states for a 
given physical situation---one only has to question the applicability of the same effective description when the strength parameter is varied
over a large range.

\begin{figure*}
\centering
{\includegraphics[width=0.8\textwidth]{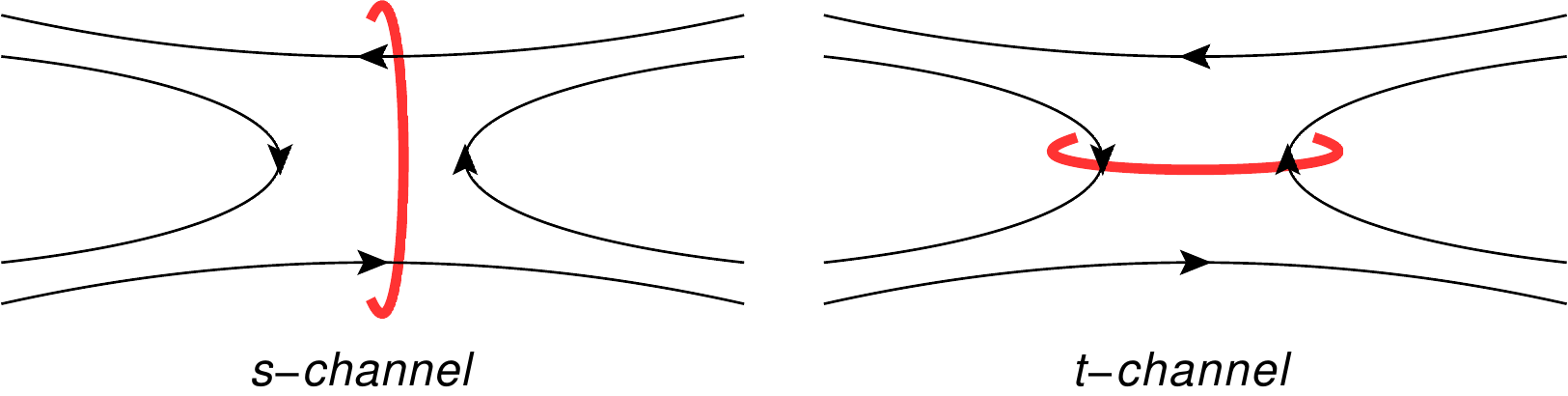}}
\caption{An illustration of the duality idea: a given quark-model diagram can either be viewed
as a $s$-channel exchange or a $t$-channel exchange depending on which interaction produces
a pole. Here the two possible clusterings are indicated by the red lines.}
\label{fig:duality}
\end{figure*}

The observation of the collective states could remind one of the
notion of duality---for a detailed review on the subject see, for example, 
Refs.~\cite{Fukugita:1976ab,Shifman:2000jv}---which can be formulated as
the observation that an infinite sum of $s$-channel poles can be matched onto an infinite sum of $t$-channel 
poles---this relation is illustrated in Fig.~\ref{fig:duality}. 
This might explain why it was possible to predict, 
for example, the existence of $X(3872)$ charmoniumlike state from simple meson exchange models
\cite{Voloshin:1976ap,DeRujula:1976zlg,Tornqvist:1991ks}.
The extraordinary states found here appear as a collective phenomenon of basically all $s$-channel pole terms included in 
the model, and one might be tempted to claim that this would still be true even if infinitely many states were
included. Then one could argue, based on the duality picture, that the very same pole could have emerged from
an infinite sum of $t$-channel poles and that it might well be more 
efficient to parametrize the binding potential by a contact term or a few $t$-channel exchanges instead of 
a large sum of $s$-channel poles. 

\section{Summary and Outlook}

Within different quark models we studied the trajectories of series of $S$-matrix poles as their coupling to a continuum channel
is increased from 0 to some large value while keeping other input parameters fixed. As outlined in the Introduction, such a study
could be understood as a naive implementation of the behavior of the QCD spectrum as the number of colors is reduced from
some very large value.

In this work we varied the confining potential, the vertex functions as well as the number of quark states included and found
that the qualitative features of the pole trajectories persisted. Although all models used were quite simple (\emph{e.g.} no vertex
dressing via $t$-channel meson exchange was involved) we still expect that the gross features described in this paper are genuine
and should appear in all models where the couplings of $s$-channel resonances to continuum states are varied over
a wide range.
Especially one is to conclude that, when unitarizing a quark model, care has to be taken with respect to the renormalization condition imposed 
for the hadronic loops. Indeed, the way how the hadronic loops are regularized within that model provides effectively a 
modeling for the quantity $\Delta$ discussed in this paper (\emph{cf.} Eq.~(\ref{piofs})) which strongly influences the
pole trajectories. Stated differently: the pole trajectories of quark model states as a function of some coupling have to be interpreted with 
caution since one gets sensitive to the regularization imposed for the hadronic loops, that is to an effect which might lie beyond the scope
of the quark model itself. As a result, the mentioned trajectories might come out as artifacts (for additional aspects about unitarizing the 
quark model we refer the reader to Ref.~\cite{Capstick:2007tv}).

Another important insight is that, when quark states couple strongly to the continuum, at least one state develops an extraordinary,
collective nature. The collectivity found here might be interpreted as the onset of effectively building a $t$-channel exchange from an infinite sum 
of $s$-channel poles. In this sense the extraordinary states might come out as hadronic molecules, at least when they are located
near relevant thresholds.

In the models studied in this paper we witnessed the appearance of one or two such extraordinary states depending on 
the renormalization condition. However, for simplicity we limited ourselves to the study of only one continuum channel. To 
gain insight into the physics of heavy mesons a natural next step is the inclusion of more channels. First 
investigations indicate that the number of collective states tends to increase in this case. 

For the selfenergies in this work we only used the expression for the scalar loop in leading order in the
meson--meson relative momentum. As a result of this the number of poles in the $S$ matrix stayed the 
same throughout the study, at least for model A. If instead
the full expression for the relativistic scalar loop had been used, there would have appeared an
additional nonanalyticity on the second sheet that leads to an additional pole.
As a result the trajectories change quantitatively. However qualitatively they stay the same and so do the conclusions 
on the ordinary and extraordinary nature of the resonances as well as on the role played by the renormalization 
scheme. Further details will be presented elsewhere.

\medskip

The authors are grateful for useful discussions with E. Eichten, J. Gegelia, F.-K. Guo, U.-G. Mei\ss ner, Yu. S. Kalashnikova, A. E. Kudryavtsev,
and A. Wirzba.
This work is supported in part by the DFG and the NSFC through funds provided to the Sino-German CRC 110 ``Symmetries and the Emergence of Structure
in QCD'' (NSFC Grant No. 11261130311). AVN acknowledges support from the Russian Science Foundation (Grant No. 15-12-30014).

\end{document}